\documentclass[iop]{emulateapj}

\usepackage{xcolor}

\shorttitle{Overturning the Case for Gravitational Powering in a \lya\ Nebula}
\shortauthors{Prescott et al.}

\newcommand{\radius}{10} 
\newcommand{\numgal}{6} 
\newcommand{\oii}{[\hbox{{\rm O}\kern 0.1em{\sc ii}}]}
\newcommand{\oiii}{[\hbox{{\rm O}\kern 0.1em{\sc iii}}]}
\newcommand{\lya}{Ly$\alpha$}

\begin{document}

\title{Overturning the Case for Gravitational Powering in the Prototypical Cooling \lya\ Nebula} 

\author{Moire K. M. Prescott\altaffilmark{1}, 
Ivelina Momcheva\altaffilmark{2}, 
Gabriel B. Brammer\altaffilmark{3}, 
Johan P. U. Fynbo\altaffilmark{1}, Palle M{\o}ller\altaffilmark{4}} 

\altaffiltext{1}{Dark Cosmology Centre, Niels Bohr Institute, University of Copenhagen, Juliane Maries Vej 30, 2100 Copenhagen {\O}, Denmark; mkmprescott@dark-cosmology.dk}
\altaffiltext{2}{Department of Astronomy, Yale University, New Haven, CT 06511, USA}
\altaffiltext{3}{Space Telescope Science Institute, 3700 San Martin Drive, Baltimore, MD 21218, USA}
\altaffiltext{4}{European Southern Observatory, Karl-Schwarzschildstrasse 2, D-85748 Garching bei M\"unchen, Germany}

\begin{abstract}

The \citet{nil06} \lya\ nebula has often been cited as the most plausible example of a \lya\ 
nebula powered by gravitational cooling.  In this paper, we bring together new data from the 
Hubble Space Telescope and the Herschel Space Observatory\footnote{Herschel is an ESA space observatory with science instruments provided by 
European-led Principal Investigator consortia and with important participation from NASA.} 
as well as comparisons to recent 
theoretical simulations in order to revisit the questions of the local environment and most 
likely power source for the \lya\ nebula.  In contrast to previous results, we find that this 
\lya\ nebula is associated with \numgal\ nearby galaxies and an obscured AGN that is offset by 
$\sim$4\arcsec$\approx$30~kpc from the \lya\ peak.  The local region is overdense relative to 
the field, by a factor of $\sim$10, and at low surface brightness levels the \lya\ emission appears 
to encircle the position of the obscured AGN, highly suggestive of a physical association.  
At the same time, we confirm that there is no compact continuum source located within 
$\sim$2-3\arcsec$\approx$15-23~kpc of the \lya\ peak.  Since the latest cold accretion simulations 
predict that the brightest \lya\ emission will be coincident with a central growing galaxy, we 
conclude that this is actually a strong argument {\it against}, rather than for, the idea that 
the nebula is gravitationally-powered.  While we may be seeing gas within cosmic filaments, this 
gas is primarily being lit up, not by gravitational energy, but due to illumination from a nearby buried AGN.

\end{abstract}


\section{Introduction}
\label{sec:intro}

There has been a long-standing debate about the power source responsible for radio-quiet Lya nebulae (or ``\lya\ blobs'').  
Unlike extended \lya\ halos observed around many radio galaxies and quasars, radio-quiet \lya\ nebulae 
do not always have an obvious single source of ionizing photons in their midst.  
Various explanations have been proposed: for example, shock-heating 
in galactic superwinds \citep{tani00, tani01, mori04}, photoionization from nearby galaxies 
and AGN \citep[e.g.,][]{cantalupo05,gea09,koll10,pres12b,overzier13}, 
and resonant scattering of centrally produced \lya\ photons \citep[e.g.,][]{moller98,laursen07,zheng11,stei11,hayes11}.  
One final explanation is that \lya\ nebulae are powered by gravitational cooling radiation within 
cold accretion streams.  This possibility has received substantial theoretical attention over the 
past few years, motivated in large part by evidence from numerical simulations that a cold mode of 
accretion could play an important if not dominant role in fueling galaxy formation 
\citep[e.g.,][]{keres05,dekel09}. 

Against this backdrop, \citet{nil06} reported the discovery of a \lya\ nebula at $z\approx3.157$ 
in the GOODS-S field.  Intriguingly, the deep GOODS imaging showed no associated continuum 
detections in any band, from the X-rays to the infrared.  The photometric redshifts available at 
the time indicated that none of the nearby galaxies were likely to be at the redshift 
of the nebula nor would they be able to provide sufficient energy to power the nebula even if they were.  
Based on the lack of any obvious 
source of ionizing photons and on the similarity between the observed \lya\ surface brightness 
profile and early predictions from cold accretion models, the authors suggested that this system could 
be powered by cooling radiation.  Since then, discussions of the range of power sources observed 
for \lya\ nebulae typically include a reference to this system as the clearest observational 
example of a ``cooling" \lya\ nebula \citep[e.g.,][]{smi08,dijk09,adams09,rasmussen09,laursen09,yang10,goerdt10,weij10,colbert11,erb11,cen13,laursen2013,tamura2013,yajima2013}.

However, a number of things have changed substantially since the initial discovery of this \lya\ nebula.  
On the observational side, grism spectroscopy, deeper X-ray catalogs, and additional imaging - spanning the 
restframe optical to far-infrared - have become available in the GOODS-S field 
thanks to the Early Release Science program \citep{windhorst2011}, 
the 2 Ms and 4 Ms Chandra Deep Field-South (CDF-S) Surveys \citep{luo2008,xue2011}, 
GOODS-Herschel \citep{elbaz11}, and HerMES \citep{levenson10,oliver12}, 
allowing for significant improvements to both photometric and spectroscopic redshift 
measurements in the region.  
In addition, some observational studies of other \lya\ nebulae have shown that the \lya\ emission can 
actually be offset substantially ($\sim$10s of kpc) from associated galaxies and obscured AGN at the same 
redshift \citep{pres09,pres12a,pres12b,pres13,yang14a}.  
On the theoretical side, simulations of ``cooling" \lya\ nebulae have been carried out with 
greater and greater sophistication over the past few years \citep[e.g.,][]{goerdt10,faucher10,rosdahl12}.  
In light of the new observational data and more advanced numerical simulations, the time is ripe 
to return to this system and take another look at its local environment and most likely power source. 
 
In this paper, we revisit the following questions: 
(1) does the \lya\ nebula have any associated galaxies, and (2) is it likely powered by cold accretion?  
We describe the datasets used in this work in Section~\ref{sec:data}, and we study the 
neighborhood of the \lya\ nebula in Section~\ref{sec:revisiting}.  
In Section~\ref{sec:powering} we discuss the most likely powering mechanism responsible for 
the \lya\ nebula, taking into account updated predictions from numerical simulations.  
Section~\ref{sec:implications} highlights some implications of this work for the larger 
class of \lya\ nebulae, and we conclude in Section~\ref{sec:conclusions}.  
We assume the standard $\Lambda$CDM cosmology ($\Omega_{M}$=0.3, $\Omega_{\Lambda}$=0.7, $h$=0.7);
the angular scale at $z=3.156$ is 7.58 kpc/\arcsec.  All magnitudes are in the AB system \citep{oke74}.

\section{Data}
\label{sec:data}

This paper builds on the large number of existing datasets and catalogs with coverage of 
the GOODS-S field.

\subsection{Ly$\alpha$ Narrowband Imaging and Spectroscopy}
The original \lya\ imaging was obtained as part of a survey for 
\lya-emitting galaxies using FORS1 on the VLT 8.2 m telescope Antu and a standard 
narrowband filter ($\lambda_c$=5055\AA; $FWHM$=59\AA, corresponding to \lya\ at $z=3.126-3.174$).  
The nebula emission was confirmed to be \lya\ at $z=3.157$ via follow-up spectroscopy from VLT/FORS1, 
which showed a single emission line with the classic asymmetric profile expected for high redshift 
\lya\ emission and no detections of the other strong lines that would have been expected from a low redshift interloper.  
Details on the \lya\ imaging and follow-up spectroscopy are given in the original survey papers \citep{nil06,nil07}.  

\subsection{HST Imaging and Grism Spectroscopy}

The \lya\ nebula is located within the GOODS-S/ERS field.  The area was originally observed as part 
of the GOODS South program in the ACS $BVRiz$ filters \citep{giav04,dickinson2004}.  
The same region was more recently targeted as part of the WFC3/ERS observations in Cycle 17 following the 
installation of WFC3 during Servicing Mission 4 \citep[GO/DD: 11359, 11360; PI: R. O'Connell][]{windhorst2011}.  
Images were acquired in the $F098M$, $F125W$ and $F160W$ filters, using two orbits in each filter.  Additional 
shallower imaging in the $F140W$ filter was done to provide a direct-image reference for a single pointing 
of ERS $G141$ grism observations (discussed below).  The $F140W$ observations partially overlap with the footprint of the 
3D-HST program (GO: 12177; PI: van Dokkum) which also provides $F140W$ images over a larger area.  The GOODS-S/ERS 
field has been observed by an array of additional ground- and space-based facilities.  
In this work, we make use of WFC3-IR $F125W$, $F140W$, and $F160W$ imaging 
and the WFC3-detected photometric catalog of the GOODS-S field compiled by \citet{skelton14}.  
The WFC3 imaging represents a significant improvement over the ground-based $J$ and $H$-band data available previously, 
as it provides high resolution imaging coverage near the critical 4000\AA\ break, 
and enabling much more robust photometric redshift estimates; see \citet{skelton14} for further information.  
We make use of the photometric redshift catalog from \citet{skelton14}, which 
is based on the code EAzY \citep{brammer08}, and morphological measurements for galaxies in the region 
taken from the CANDELS morphological catalog \citep{vanderwel2012}.  

The  \lya\ nebula is serendipitously located in the one and only grism pointing observed as part of 
the ERS program \citep[GO/DD: 11359, 11360; PI: R. O'Connell]{windhorst2011} 
The observations were done in both the $G102$ and $G141$ grisms, spanning $0.8-1.2\; \mu m$ and $1.1-1.7\; \mu m$, 
respectively.  The two grism provide resolutions of $R\sim210$ and 130.  The pointing was observed for two orbits 
with each grism.  In this work we use only the $G141$ grism observations, which covers the \oii$\lambda$3727 emission 
line out to $z\approx3.2$, as none of the objects relevant to this work have significant detections in the $G102$ spectra.  
The grism data were reduced by the 3D-HST team in a manner similar to that described in Brammer et al. (2012).  
Further details on the extraction and redshift-fitting methods will be presented in Momcheva et al. (in prep.).

\subsection{Mid- and Far-Infrared Data}

The original paper presented Spitzer IRAC and MIPS imaging of the field \citep[][their Figure~2]{nil06}.  
At longer wavelengths, Herschel PACS and SPIRE coverage of the field now 
exists from GOODS-Herschel and HerMES \citep{elbaz11,oliver12}.  
In addition, improvements have been made in extracting accurate IRAC/MIPS photometry properly 
accounting for source blending and the complex PSF of these instruments, and 
there has been extensive work on the best way to use IRAC/MIPS colors to distinguish 
AGN from star-forming galaxies \citep{stern05,lac04,lac07,don08,don12}.

In this work, the mid- and far-infrared photometry are taken from the GOODS-Herschel catalog, which includes 
measurements from the GOODS-Spitzer Legacy Survey (Spitzer/IRAC+MIPS; P.I. Mark Dickinson) 
and the GOODS-Herschel Survey \citep[Herschel/PACS;][]{elbaz11}, accessed via 
HeDaM.\footnote{http://hedam.lam.fr}  
Source positions in the GOODS-Herschel catalog were determined first using a weighted average of the 
3.6 and 4.5 micron images, and then empirical PSFs at these source locations 
were fit to the mid- and far-infrared maps in order to determine the flux for 
each object and account for source blending.  Further details are given 
in the GOODS-Herschel release documentation.\footnote{http://hedam.lam.fr/GOODS-Herschel/goodss-data.php}

The field has been observed in the far-infrared with Herschel/SPIRE as part of HerMES
\citep{oliver12,levenson10,roseboom2010,roseboom2012}.  In this work, we make use of Herschel/SPIRE photometric 
measurements from the MIPS 24$\mu$m-matched catalog developed by the HerMES Team (HerMES Team, private communication).

\subsection{Chandra X-ray Coverage}

Since the original paper, the Chandra X-ray coverage of GOODS-S has improved substantially 
from the original Chandra Deep Field-South 1 Ms survey, first by a factor of two and then 
by a factor of four in total exposure time.  
The resulting Chandra Deep Field-South 2 Ms and 4 Ms survey catalogs \citep{luo2008,xue2011} 
provide a sample 740 X-ray detected sources down to limiting on-axis fluxes of 
$\approx$3.2$\times$10$^{-17}$, 9.1$\times$10$^{-18}$, and 5.5$\times$10$^{-17}$ erg s$^{-1}$ cm$^{-2}$ 
for the 0.5-8 keV full band, the 0.5-2 keV soft band, and the 2-8 keV hard band, respectively.

\section{Revisiting the Neighborhood}
\label{sec:revisiting}

\citet{nil06} listed 8 objects within a 10\arcsec\ radius that had detections 
in at least 8 photometric bands, and an additional 8 nearby sources that were detected 
in only a subset of those bands, too few for reliable photometric redshifts to be estimated.  
Of the 8 sources with derived photometric redshifts, they presented projected distances and 
photometric redshifts in their Table~2.  None appeared to be precisely at the \lya\ nebula redshift, 
although four of these sources were consistent with it to within the relatively large quoted errors 
(Sources 3, 5, 6, and 8).  Two of these sources were not discussed further (Sources 5 and 8).  
The authors argued that Source 3 was unlikely to power the nebula, even if it were at the 
correct redshift, because it did not appear to emit enough ionizing photons.  
Finally, in the case of Source 6, a mid-infrared source that was detected in the IRAC and MIPS 
bands, they concluded the object was most 
likely an unassociated starburst galaxy at high redshift ($z\approx5.5$) 
based on comparisons to Spitzer IRAC/MIPS color-color diagnostic diagrams from \citet{ivison04}. 

Using the new and improved datasets described in the previous section, 
we revisit the following question: is this \lya\ nebula truly a line-emitting 
nebula without any associated galaxies?

\subsection{Searching for Associated Galaxies}

\subsubsection{Photometric Redshifts}

\begin{figure}[t]
\center
\includegraphics[angle=0,width=7in]{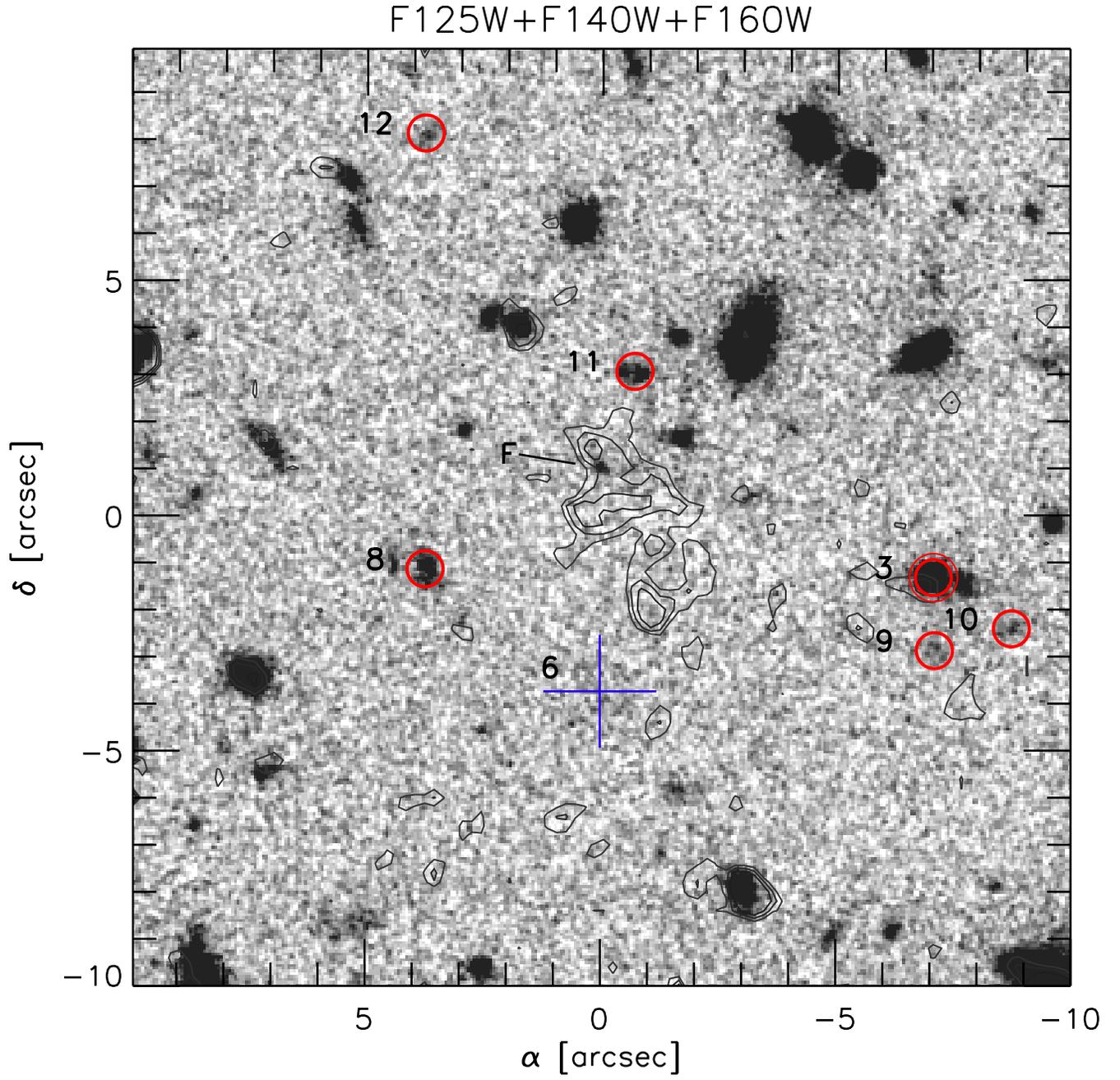} 
\caption[]{
Stacked WFC3-IR/$F125W$, $F140W$, and $F160W$ imaging of the region around the \lya\ nebula.  
The \lya\ emission is shown with black contours at \lya\ surface brightness 
levels of [5.3,7.0,8.9]$\times10^{-18}$ erg s$^{-1}$ cm$^{-2}$ arcsec$^{-2}$ 
after smoothing by $0.4\times0.4$\arcsec.  
Sources with photometric redshifts within $\Delta z=0.15$ are shown (red circles), 
while Source 3, the galaxy with a consistent grism redshift, is indicated with a double red circle.
The IRAC centroid of Source 6 from the GOODS-Herschel catalog 
is indicated with a blue cross.  
The galaxy located closest to the \lya\ peak, labeled 
`F', is a foreground galaxy at $z\approx1$. 
}
\label{fig:masterstack}
\end{figure}

\begin{figure}[t]
\center
\includegraphics[angle=0,width=7in]{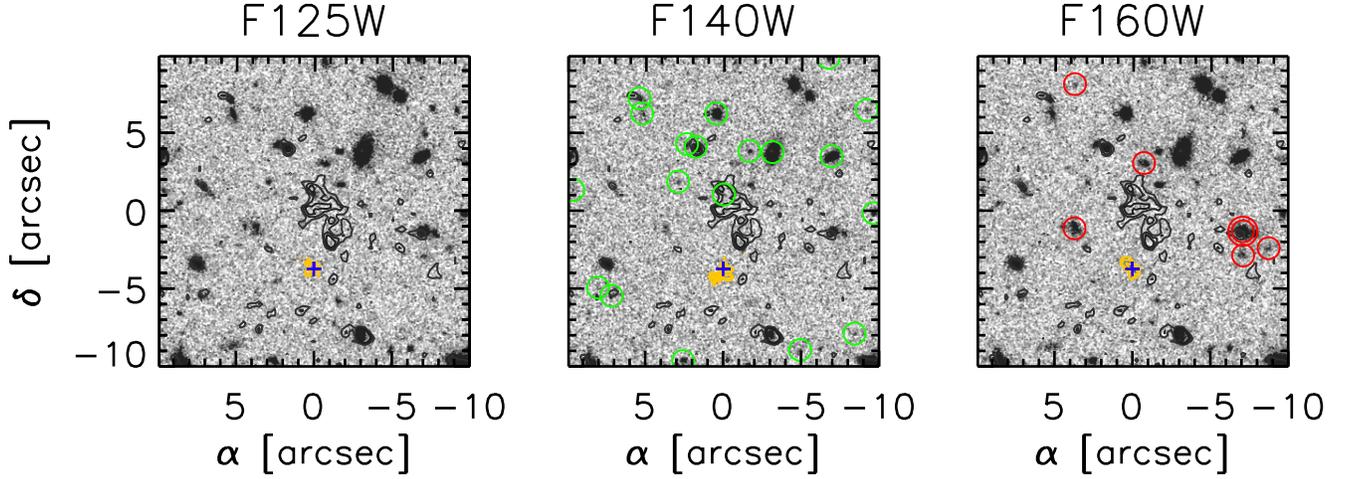} 
\caption[]{
Region around the \lya\ nebula in the individual WFC3-IR/$F125W$, $F140W$, and $F160W$ images.  
In all panels the \lya\ nebula is shown with black contours at \lya\ surface brightness 
levels of [5.3,7.0,8.9]$\times10^{-18}$ erg s$^{-1}$ cm$^{-2}$ arcsec$^{-2}$ 
after smoothing by $0.4\times0.4$\arcsec.  
The IRAC centroid of Source 6 from the GOODS-Herschel catalog is indicated with a blue cross, and the yellow contours 
in each panel highlight diffuse emission in the corresponding WFC3-IR image 
that is detected within $R<1.3$\arcsec\ of Source 6.  
{\bf Left:} The WFC3 $F125W$ image.  
The $F125W$ surface brightness contours at the position of Source 6 (yellow lines) correspond to 
$SB_{F125W}=[0.4,0.6,0.8]\times10^{-20}$ erg s$^{-1}$ cm$^{-2}$ \AA$^{-1}$ arcsec$^{-2}$.  
{\bf Middle:} The WFC3 $F140W$ image; foreground sources with photometric redshifts within 
$\Delta z\pm0.15$ of $z=1$ are indicated (green circles).  
The $F140W$ surface brightness contours at the position of Source 6 (yellow lines) correspond to 
$SB_{F140W}=[0.8,1.0,1.2]\times10^{-20}$ erg s$^{-1}$ cm$^{-2}$ \AA$^{-1}$ arcsec$^{-2}$.  
{\bf Right:} The WFC3 $F160W$ image; sources with photometric redshifts 
within $\Delta z\pm0.15$ of the redshift of the \lya\ nebula 
are indicated (red circles).  
The source with a detection of [OII] at $z=3.153$ in the 
$G141$ grism data is shown with a larger red circle.
The $F160W$ surface brightness contours at the position of Source 6 (yellow lines) correspond to 
$SB_{F160W}=[0.8,1.0,1.2]\times10^{-20}$ erg s$^{-1}$ cm$^{-2}$ \AA$^{-1}$ arcsec$^{-2}$.  
}
\label{fig:spatialdistribution}
\end{figure}

\begin{figure}[th]
\center
\includegraphics[angle=0,width=7in]{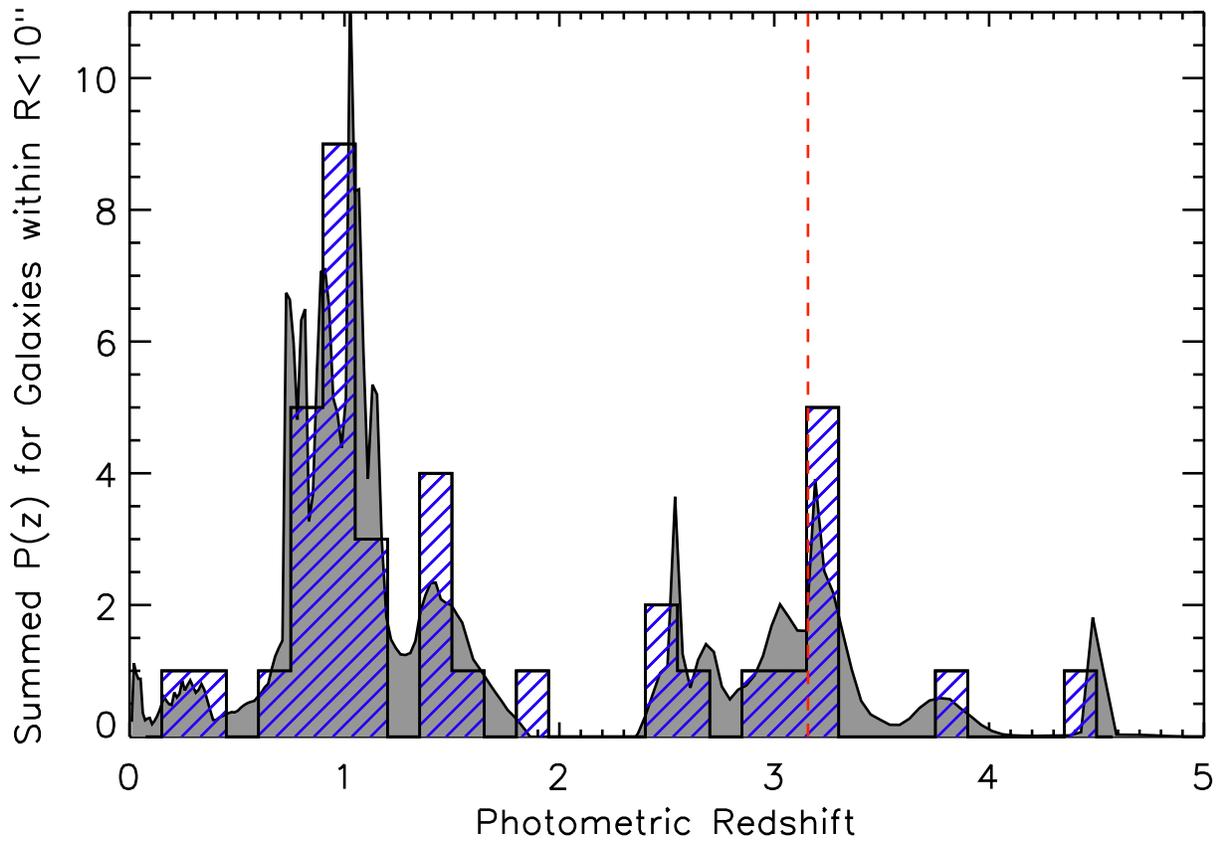} 
\caption[]{Photometric redshift distribution for sources within $R<\radius\arcsec$ of the \lya\ nebula.  
The summed P(z) distribution is shown (grey shading) along with 
a histogram of the best fit photometric redshifts (blue hashed histogram).  
The redshift of the \lya\ nebula is indicated with the red dashed line.
}
\label{fig:zdist}
\end{figure}

Figure~\ref{fig:masterstack} shows a stack of the WFC3-IR/$F125W$, $F140W$, and $F160W$ imaging in the vicinity of 
the \lya\ nebula, shown as contours,\footnote{We note that the surface brightness levels for the 
contours plotted in \citet{nil06} appear to have been stated incorrectly in the caption of their Figure~1.} 
while Figure~\ref{fig:spatialdistribution} shows the individual bands.
Using the 3D-HST/CANDELS photometric redshift catalog \citep{skelton14}, we select sources 
that lie within a \radius\arcsec\ radius of the \lya\ nebula.  
Figure~\ref{fig:zdist} shows the distribution of photometric redshifts in 
this region, both as a histogram and as a sum of the individual galaxy P(z) distributions.  
The photometric redshift distribution shows two main peaks: one at $z\approx1$ and one around $z\approx3.1$, i.e., roughly 
the redshift of the nebula.  
Formally, \numgal\ sources within this local region ($R<10$\arcsec) have photometric redshifts 
within $\Delta z\pm0.15$ of the nebula redshift; the positions and 
photometric redshifts of these ``photometric redshift members'' are given in 
Table~\ref{tab:galmember}.  
We plot the selected photometric redshift members on the HST imaging (Figure~\ref{fig:spatialdistribution}, righthand panel).  
Of these \numgal\ sources, photometric redshift estimates for two 
(Source 3 and 8) were given in the original paper; the new photometric redshifts for these sources are consistent with the previous estimates, 
given the quoted error bars.  The photometric redshifts of the remaining sources were previously unknown.  

We note that while photometric redshifts will clearly be less accurate and more susceptible to biases, 
especially at fainter magnitudes, relative to true spectroscopic redshifts, a recent study of a suite of photometric redshift codes 
found that individual codes based on EAzY \citep{brammer08} do reasonably well in terms of their reported errors 
(i.e., corresponding to the width of the reported P(z) peak), underestimating by about 5-15\% \citep{dahlen13}.  
The scatter ($\sigma_{0} =$ rms[$\Delta$z/(1+z)]) between the spectroscopic redshift and the median photometric redshift 
was found to increase from $\sigma_{0}=0.04$ at $m_{H}<24$ mag to $0.09$ at $m_{H}\sim26$ mag, while the outlier fraction 
increased from about 10\% to 20\%.  Thus by using a redshift window of $\Delta z\pm0.15$, we can be reasonably confident 
that we are primarily selecting galaxies at the redshift of the nebula even at these fainter magnitudes.

To test the significance of the observed photometric redshift peak coincident with the nebula, we redid the analysis for 
10000 random apertures drawn from the entire GOODS-S field.  We find that at $z\sim3$, it is not common to find a peak 
in the photometric redshift distribution that is as strong as what is seen at the position of the \lya\ nebula: 
formally there is a 
2.0\% 
chance of randomly selecting a region with comparable area under the summed P(z) distribution, 
or a 
0.4\% 
chance of randomly selecting a region with at least \numgal\ galaxies with best fit photometric redshifts 
within $\Delta z\pm0.15$ of the nebula redshift.

\begin{figure}[th]
\center
\includegraphics[angle=0,width=7in]{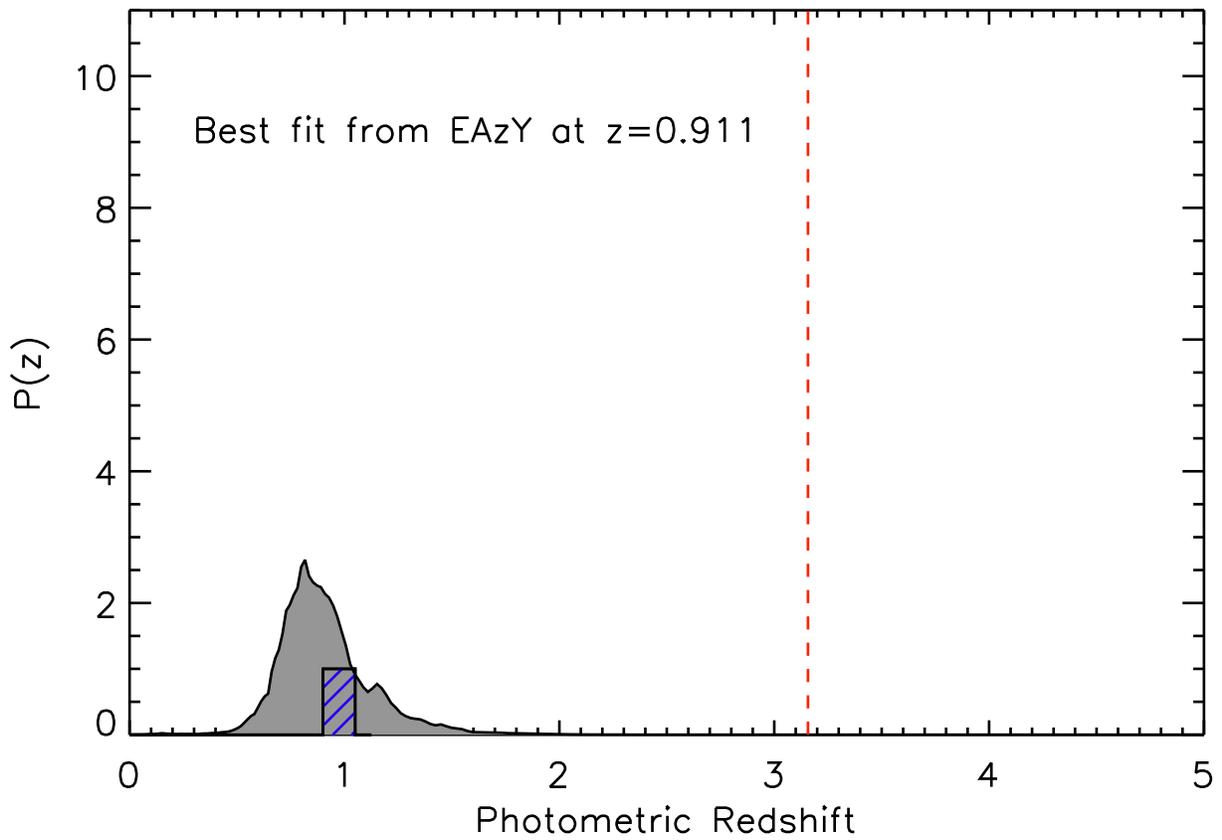} 
\caption[]{Photometric redshift distribution for Galaxy `F', the faint galaxy closest to the peak of the \lya\ emission. 
The P(z) distribution is shown (grey shading) along with a 
histogram showing the best fit photometric redshift (blue hashed).  
The redshift of the \lya\ nebula is indicated with the red dashed line.
}
\label{fig:zdist41245}
\end{figure}

\begin{figure}[th]
\center
\includegraphics[angle=0,width=7in]{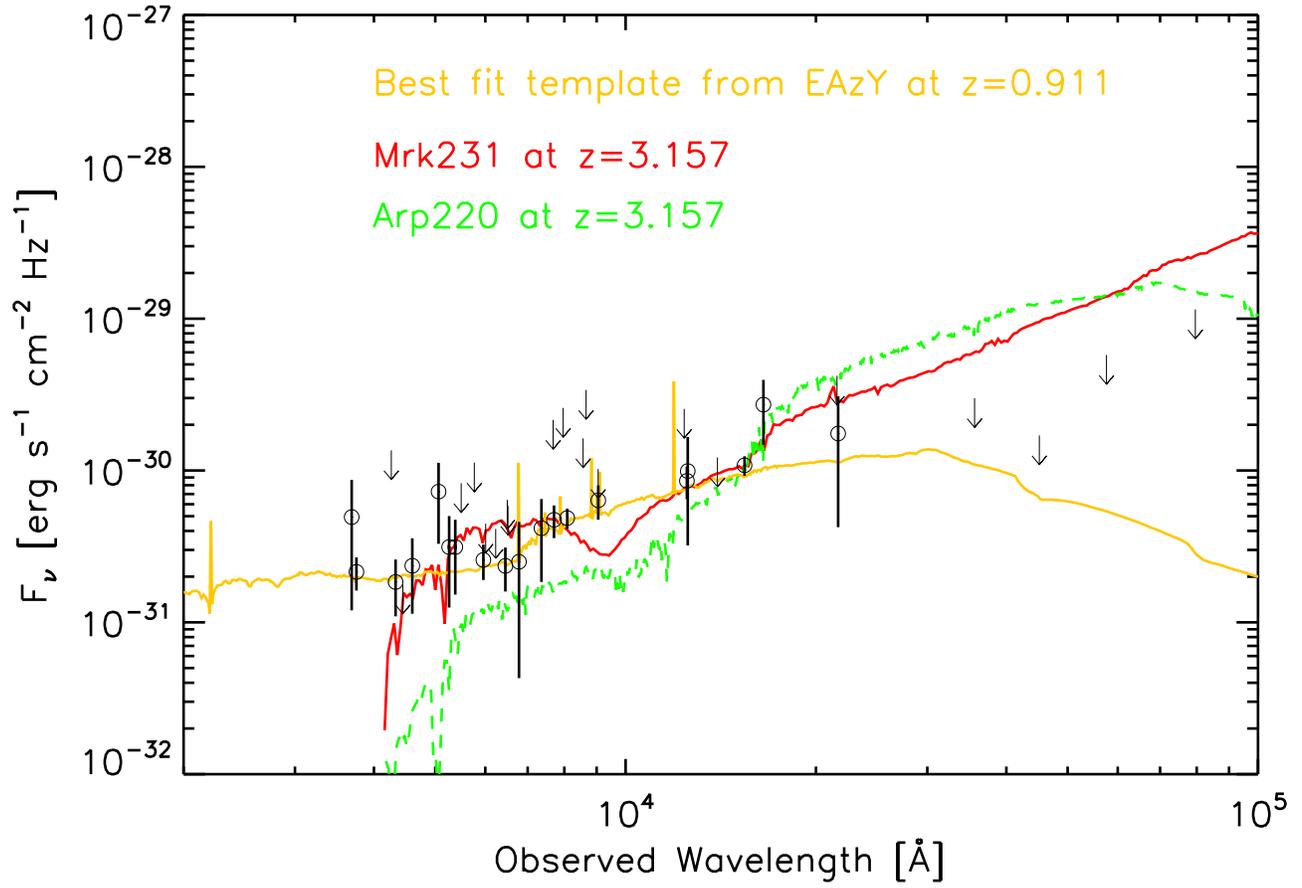} 
\caption[]{
SED (open black circles and arrows representing $3\sigma$ upper limits) 
and best-fit EAzY template (yellow solid line) 
for Galaxy `F', the faint source closest to the peak of the \lya\ emission, as compared 
with Mrk231 (red solid line) and Arp220 (green dashed line) templates shifted to $z=3.157$.  
Despite being located very near the \lya\ peak, this source is inconsistent with 
being at the nebula redshift.  It is instead part of the overdensity of $z\approx1$ sources 
that overlaps this location.
}
\label{fig:sed41245}
\end{figure}

Interestingly, the very faint source visible in the $F160W$ image that is located closest to 
the peak of the \lya\ emission turns out to be one of the $z\approx1$ sources (labeled as Galaxy 
`F' in Figure~\ref{fig:masterstack}).  The measured 
SED and P(z) derived by EAzY for this source show that it is well-described by the EAzY fit at $z=0.91$ 
with a stellar mass of $\approx1.2\times10^{8}$ M$_{\odot}$, 
and that it is inconsistent with either an AGN or starburst template at the nebula redshift 
(Figures~\ref{fig:zdist41245}-\ref{fig:sed41245}).  Thus, we confirm that there does not appear to be 
any associated continuum source within $\sim$2-3\arcsec$\approx$15-23~kpc of the 
peak of the \lya\ emission.

We note that gravitational lensing from the foreground sources at $z\approx1$ should only have a minimal 
effect on the nebula.  If we take all galaxies in the local vicinity that have photometric redshifts within $\pm0.15$ of 
$z=1$, the total stellar mass derived from the SED fitting is $\approx3\times10^{10}$ M$_{\odot}$, with the 
brighter (and presumably more massive) $z\approx1$ sources being located about 5\arcsec\ to the northwest of the nebula 
(Figure~\ref{fig:spatialdistribution}, middle panel).  If the corresponding total mass of the foreground structure 
is of order $3\times10^{11}$ M$_{\odot}$, the expected Einstein radius would be $\approx$0.2\arcsec, 
insignificant compared to the full extent of the \lya\ emission, and the magnification at the position of the \lya\ peak 
would only be about 3\%.  If we consider only the source located closest to the center 
(Galaxy `F', with a stellar mass of $1.2\times10^{8}$ M$_{\odot}$), the expended Einstein radius 
would be $\approx$0.05\arcsec.  
Thus, neither the large scale morphology nor the surface brightness of the nebula is 
significantly affected by gravitational lensing from the $z\approx1$ sources.

\subsubsection{Grism Redshifts}

\begin{figure}[th]
\center
\includegraphics[angle=0,width=7in]{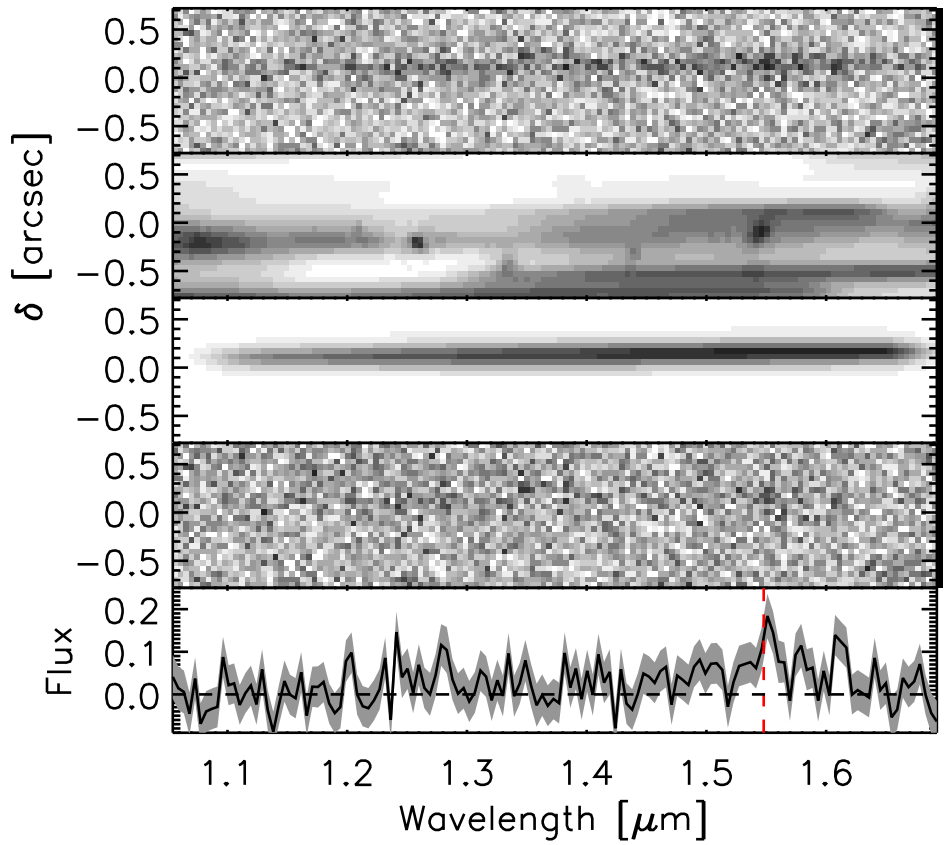} 
\caption[]{
Grism spectroscopy of Source 3, showing the 2D grism spectrum (top row), the contamination model corresponding 
to flux from nearby sources (second row), the continuum model (third row), the continuum- and contamination-subtracted 
grism spectrum (fourth row), and the 1D spectral extraction (bottom row) with the corresponding errors overplotted 
(grey shading) and the position of the measured [OII] line indicated (vertical dashed red line).
}
\label{fig:source3}
\end{figure}

We next searched the grism redshift catalog for sources with grism 
redshifts consistent with the nebula.  We find that one of 
the \numgal\ photometric redshift neighbors (Source 3) shows a weak 
[OII] detection (Figure~\ref{fig:source3}).  This corresponds to a measured grism redshift of 
$z_{grism}=3.153_{-0.004}^{+0.019}$ 
(Table~\ref{tab:galmember}), consistent with the photometric redshift ($z_{phot}=3.206^{+0.065}_{-0.058}$) 
and with the nebula redshift measured from the \lya\ emission line ($z_{Ly\alpha}=3.157$).
Neither the extended nebula itself nor any of the other galaxy members, which are all much fainter in $F160W$ continuum magnitude than Source 3, show a line detection in either the $G102$ or $G141$ grism spectroscopy.

\subsubsection{Source 6 - The IRAC/MIPS Source}
\label{sec:source6}

\begin{figure}[th]
\center
\includegraphics[angle=0,width=6in]{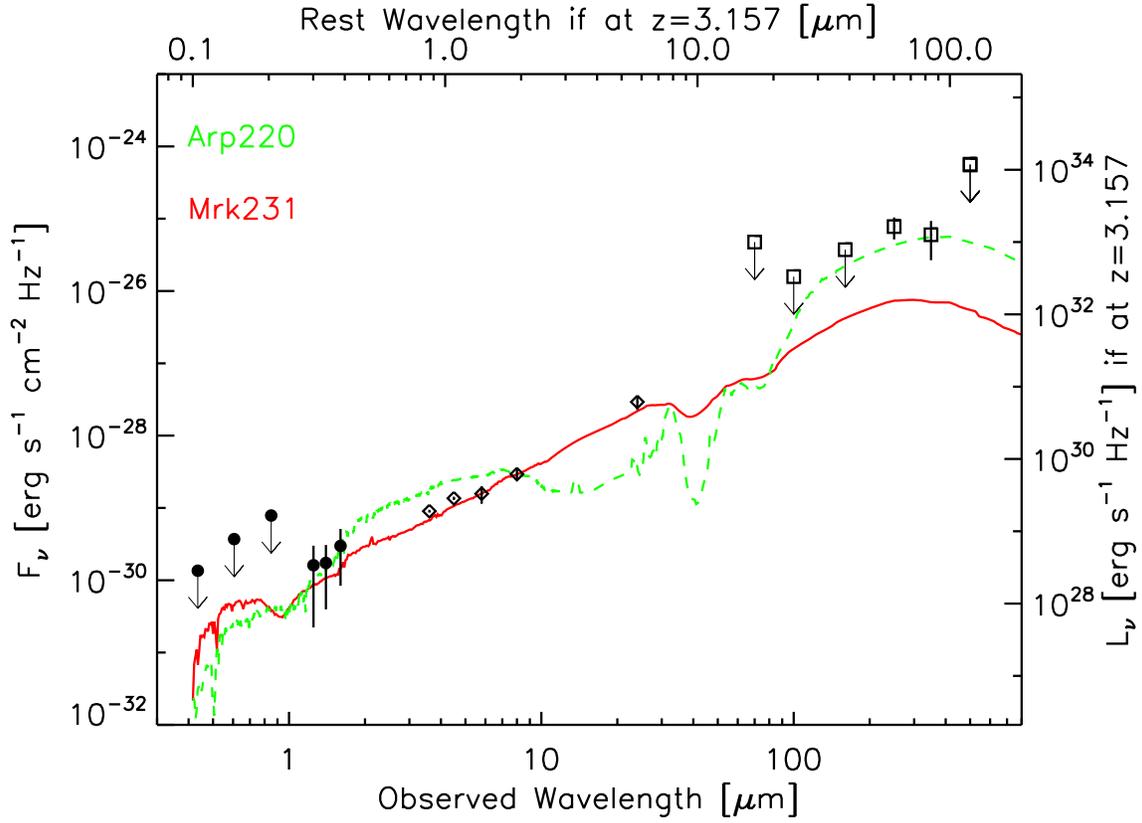} 
\caption[]{
Source 6 SED from HST (filled circles), Spitzer (open diamonds), and Herschel (open squares).  
All upper limits are shown at 3$\sigma$.  
Mrk231 (red solid line) and Arp220 (green dashed line) templates are shown for reference \citep{polletta07}, 
scaled to the 8\micron\ data point.
The top axis gives the corresponding rest wavelength and the right axis the corresponding luminosity 
density L$_{\nu}$, assuming Source 6 is at the redshift of the \lya\ nebula.
The mid-infrared data are much better fit by the Mrk231 (AGN) template; 
the Herschel detections may be contaminated by source confusion, but suggest there may 
be an additional cool dust component.  
}
\label{fig:source6}
\end{figure}

We take a second look at Source 6, the mid-infrared source first identified in the 
Spitzer IRAC and MIPS data.  We plot the full SED including the original 
GOODS-S imaging and Spitzer data, as well as the new HST/WFC3-IR imaging and 
measurements from Herschel/PACS and SPIRE (Figure~\ref{fig:source6}).  
The source shows diffuse emission in the WFC3-IR imaging, but it is below the detection 
limit in the combined WFC3 detection image used in \citet{skelton14}; 
therefore, we measure the photometry in these bands 
using a 2.4\arcsec\ radius aperture at the position of Source 6, as determined from 
the IRAC 3.6 and 4.5$\mu$m imaging.  The IRAC/MIPS photometry and Herschel PACS upper 
limits are taken from the GOODS-Herschel catalog \citep{elbaz11}, and the Herschel SPIRE 
measurements are from HerMES (HerMES Team, private communication).  
We note that Source 6 is not detected in either the WFC3/$G102$ or $G141$ grism data.

Using these updated measurements, we derive the following MIR colors (relevant for the \citet{ivison04} 
color-color selection): $S_{24.0}/S_{8.0} = 9.959\pm2.792$ and $S_{8.0}/S_{4.5} = 2.160\pm0.380$.  
While \citet{nil06} suggest Source 6 is an obscured starburst at an approximate redshift of $z\approx5.5$ 
based on its MIR colors, the MIR color measurements taken from the GOODS-Herschel catalog place Source 6 
elsewhere in the \citet{ivison04} color space, much higher in $S_{24.0}/S_{8.0}$ and 
therefore much closer to the 
Mrk231 (i.e., AGN) locus.  It is possible that the original measurements did not adequately account 
for source blending and the complex MIPS PSF, yielding an incorrect 24$\mu$m flux for this relatively 
faint MIR source that happens to lie near the Airy ring from a brighter neighboring galaxy.

Furthermore, AGN selection based on MIR colors has been explored in great detail since the time of the 
original paper \citep{stern05,lac04,lac07,don08,don12}.  The measured MIR colors relevant for these 
color-color diagnostics are: $log(S_{8.0}/S_{4.5}) = 0.335^{+0.070}_{-0.084}$ and 
$log(S_{5.8}/S_{3.6}) = 0.242^{+0.107}_{-0.142}$.  These measurements place Source 6 solidly 
within the AGN color selection window of \citet{lac04,lac07}, 
and within the more recent IRAC power-law AGN selection box of \citet{don12}.  
From the mid-infrared colors, Source 6 is therefore very likely an obscured AGN and 
not a starburst galaxy.  The Herschel detections are susceptible to source confusion 
but suggest that there may be an additional cool dust component, e.g., emission from 
the host galaxy.  

It is true that Source 6 was not detected in the deep 1 Ms CDF-S X-ray imaging, 
as pointed out in the original paper.  We find that the source is still not detected in the more 
recent 2 Ms and 4 Ms CDF-S catalogs \citep{luo2008,xue2011}.  However, a study of power-law IRAC 
selected AGN candidates in the 2 Ms Chandra Deep Field-North (CDF-N) X-ray imaging showed 
23\% (35\%) were not detected above 3$\sigma$ (5$\sigma$) in any X-ray band \citep{donley07}, 
while stacking of the X-ray non-detected sources still showed evidence for hard emission consistent with 
AGN rather than star formation powering \citep{don12}.
Thus, it appears that the lack of an X-ray detection from Source 6, even in deep Chandra data, cannot 
necessarily be used to rule out the presence of an intrinsically luminous AGN that is simply heavily 
obscured at X-ray wavelengths. 

Based on the photometry derived from the original GOODS-S HST and Spitzer/IRAC data and the 
new ERS imaging, 
we use the photometric redshift code EAzY \citep{brammer08} to estimate a photometric redshift for Source 6.  
Including the IRAC photometric points and the standard EAzY star-forming galaxy templates results in a 
redshift of $z\approx[3.2-5.6]$, similar to what was found in the original paper.  
However, if Source 6 is an obscured AGN, as we have argued, the IRAC bands will be 
dominated by AGN emission, making a photometric redshift estimate based on only star-forming 
templates highly unreliable.  The photometric redshift estimate will be skewed to higher redshifts ($z\gtrsim4$) 
by the need to fit the bright 8.0$\mu$m point with the brightest part of the star-forming SED, i.e., 
the peak located at restframe $\sim$1.6$\mu$m.  
Adding the Mrk~231 AGN template to EAzY and including both the IRAC and MIPS detections (the Herschel detections and upper 
limits being excluded due to possible source confusion) yields a slightly lower redshift range of $z\approx[2.8,5.2]$ that 
straddles the redshift of the \lya\ nebula.  This estimate is not very robust, since fitting the power-law MIR SED 
with an AGN template does not contribute a lot of discriminatory power, but it illustrates that $z\approx3.2$ is a plausible 
redshift for Source 6.

\subsection{Spatial Distribution of the Associated Galaxies and the Larger Scale \lya\ Emission}
\label{sec:spatial}

\begin{figure}[t]
\center
\includegraphics[angle=0,width=7in]{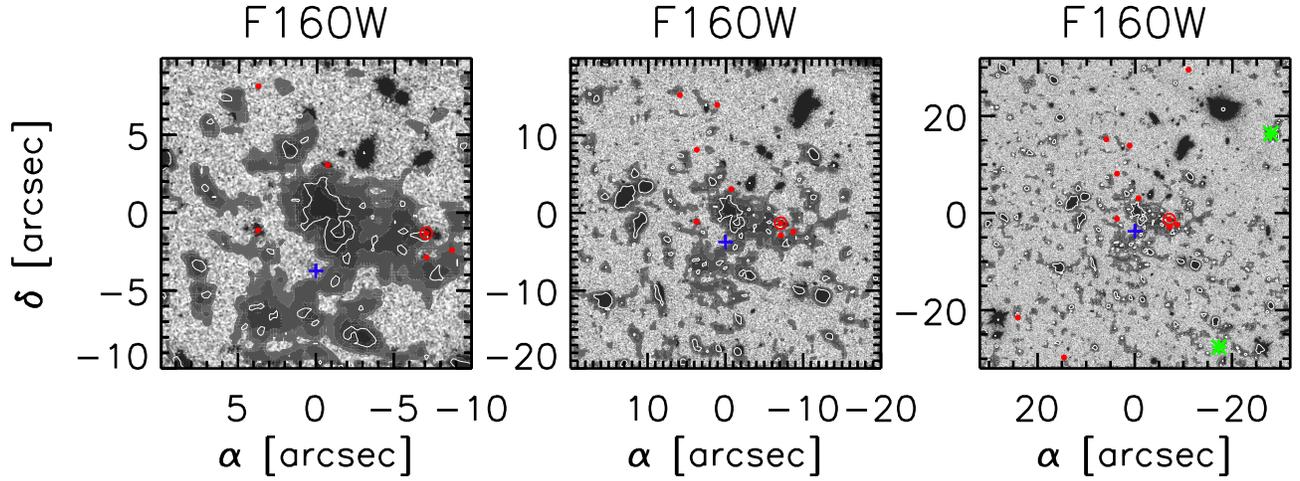} 
\caption[]{
Spatial extent of the \lya\ emission overplotted on the $F160W$ imaging.  
In all panels the \lya\ nebula, as shown in previous figures, is outlined with white contours (corresponding to 
a surface brightness level of 5.3$\times10^{-18}$ erg s$^{-1}$ cm$^{-2}$ arcsec$^{-2}$), 
while additional \lya\ surface brightness contours are shown in greyscale 
at [1.3,1.9,3.2,4.4,6.3]$\times10^{-18}$ erg s$^{-1}$ cm$^{-2}$ arcsec$^{-2}$ after 
smoothing by $0.4\times0.4$\arcsec.  
Sources with photometric redshifts 
within $\Delta z\pm0.15$ of the redshift of the \lya\ nebula are highlighted with red points, and 
Source 3, the source with a weak [OII] detection at $z=3.153$ in the $G141$ grism data, is shown with a 
larger red circle.  The position of Source 6 is indicated with a blue cross.  LAEs from 
\citet{nil07} are shown as green stars. 
}
\label{fig:contours}
\end{figure}

The spatial distribution of likely members relative to the \lya\ nebula is shown in 
Figure~\ref{fig:spatialdistribution}.  
The \lya\ nebula lies between \numgal\ galaxies with 
photometric redshifts consistent with being at the nebula redshift, one of which has a grism 
redshift as well.  
In Figure~\ref{fig:contours}, we show the full extent of the \lya\ emission in the original \lya\ imaging, within three spatial windows 
(10\arcsec$\times$10\arcsec, 20\arcsec$\times$20\arcsec, and 32\arcsec$\times$32\arcsec). 
At lower surface brightnesses than shown in the discovery paper, it appears that there is a nearly complete 
bridge of \lya\ emission between the peak of the nebula, one associated galaxy to the north, and 
three associated galaxies to the west (including Source 3, the source with a consistent grism redshift).  
At even lower surface brightnesses, the diffuse \lya\ emission connects all but one of the member galaxies 
and traces out what appears to be two intersecting filaments, a partial ring, or an asymmetric biconical nebula 
roughly centered on the position of Source 6.  
These morphological results and the result that Source 6 is an obscured AGN plausibly at the redshift of the nebula (Section~\ref{sec:source6}) 
are together highly suggestive that Source 6 is indeed directly associated with the presence of the \lya\ nebula.  
Further observations, e.g., detecting CO emission lines at submillimeter wavelengths, will allow us to confirm the redshift of Source 
6 and verify that we are seeing a \lya\ nebula surrounding an obscured AGN.

\subsection{The Local Environment} 

\begin{figure}[t]
\center
\includegraphics[angle=0,width=4.5in]{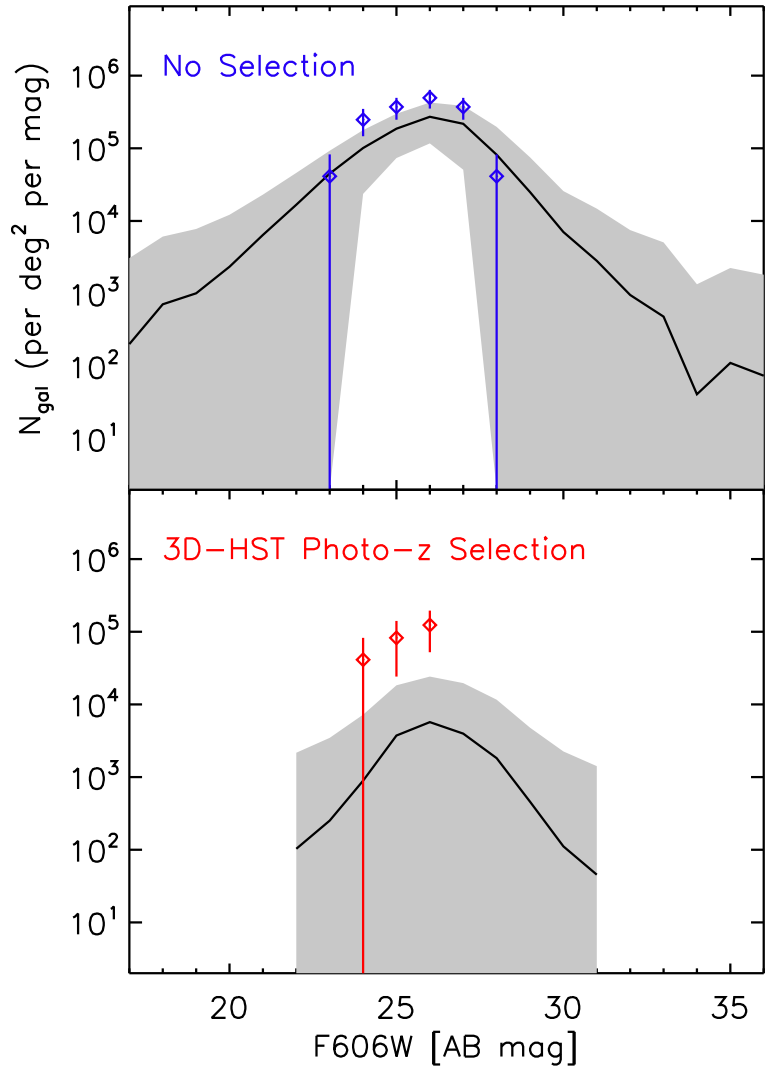} 
\caption[]{
{\bf Top:} Number counts in the $F606W$ band 
measured from the region around the \lya\ nebula ($R<\radius\arcsec$; blue diamonds) 
and from within random $R<\radius\arcsec$ apertures across the entire GOODS-S field 
(with the mean and standard deviation shown as the black line and grey region, respectively).  
The $F606W$ band was chosen to be consistent with previous studies of the 
environments of \lya\ nebulae at $z\approx3$ \citep{pres12b}.  
{\bf Bottom:} Sources with photometric redshifts within $\Delta z\pm0.15$ 
of the \lya\ nebula redshift measured from the region around the \lya\ nebula ($R<\radius\arcsec$; red diamonds) 
and from within random $R<\radius\arcsec$ apertures across the entire GOODS-S field 
(with the mean and standard deviation shown as the black line and grey region, respectively).  
}
\label{fig:allcounts}
\end{figure}

Thus far we have shown that there are in fact galaxies associated with the \lya\ nebula.  
Here we assess whether the region is in fact overdense relative to the field.  
To be consistent with previous studies of the environments of \lya\ nebulae at $z\approx3$ 
\citep{pres12b}, 
we start by plotting the number counts in the F606W band measured from within the $R<10$\arcsec\ region around the \lya\ nebula 
and from random $R<10$\arcsec\ apertures across the full GOODS-S field (Figure~\ref{fig:allcounts}; top panel).  
From this naive approach, there is only mild evidence for an enhancement in the region of the nebula; 
the number counts are slightly higher than in the field at $m_{F606W}\sim25$ but consistent to 
within the errors, i.e., the scatter measured from the random aperture test.  In addition, 
we have also seen from the photometric redshift analysis that 
there is an overdensity of $z\approx1$ sources overlapping this 
region, which will necessarily boost the raw number counts.
Next we look at the number density of sources with photometric redshifts within $\Delta z\pm0.15$ of 
the redshift of the nebula ($z=3.157$).  In the bottom panel of Figure~\ref{fig:allcounts} we plot the corresponding 
densities for the region of the nebula and for random apertures drawn from the full GOODS-S field.  
The region of the nebula shows a factor of $\gtrsim$10 overdensity at $m_{F606W}\approx25$ relative to 
the field galaxy population at the same photometric redshift, even when accounting for the scatter measured 
from randomly placed apertures.  Thus, not only are there galaxies associated with the \lya\ nebula, it is in fact 
sitting in a galaxy-overdense region of the Universe.

\subsection{Properties of the Associated Galaxies}

\begin{figure}[t]
\center
\includegraphics[angle=0,width=7in]{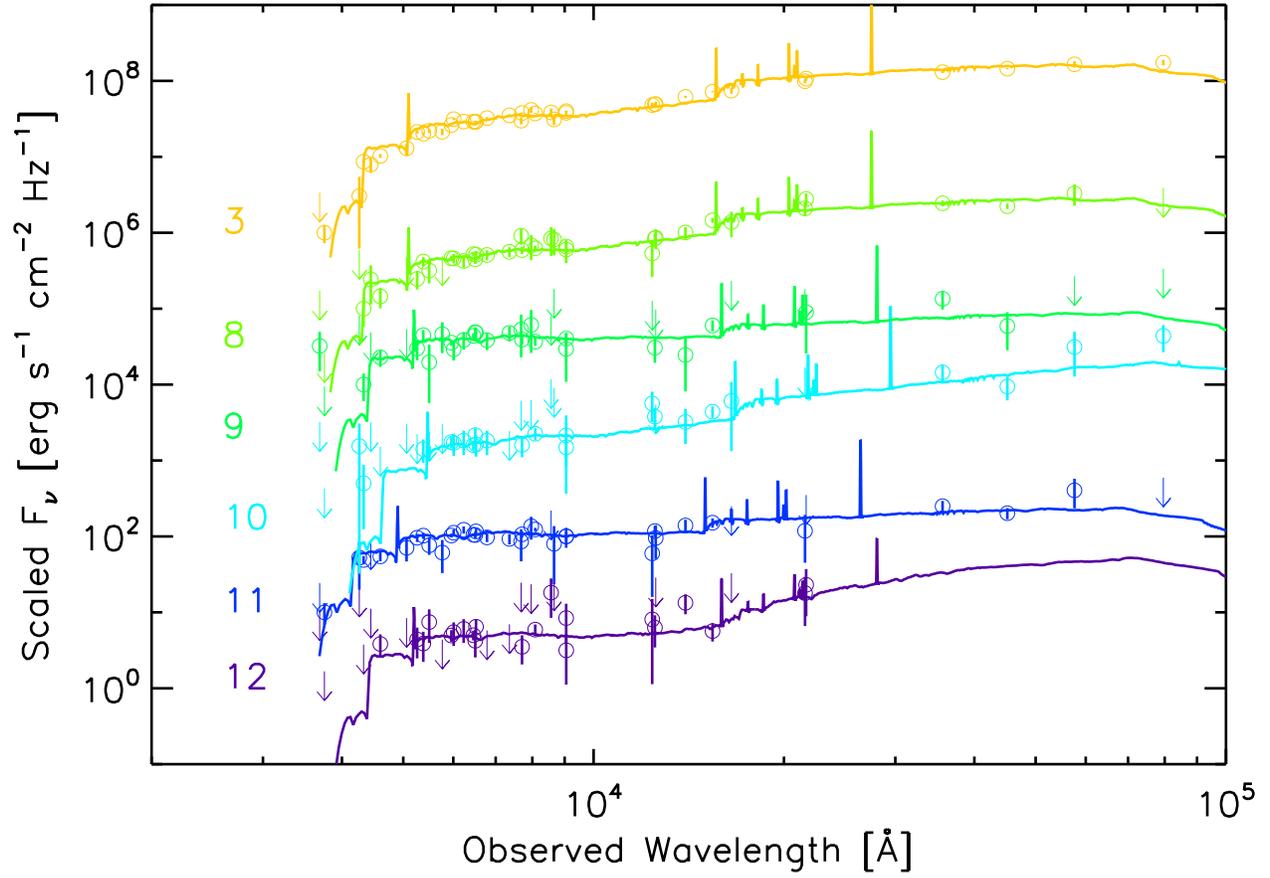} 
\caption[]{
Photometry and best-fit EAzY SEDs (open circles and arrows representing $3\sigma$ 
upper limits) of the \numgal\ galaxies likely associated with the 
\lya\ nebula \citep{skelton14}.  
Numbers correspond to the source IDs in Table~\ref{tab:galmember}.
}
\label{fig:membersedall}
\end{figure}

\begin{figure}[th]
\center
\includegraphics[angle=0,width=5.5in]{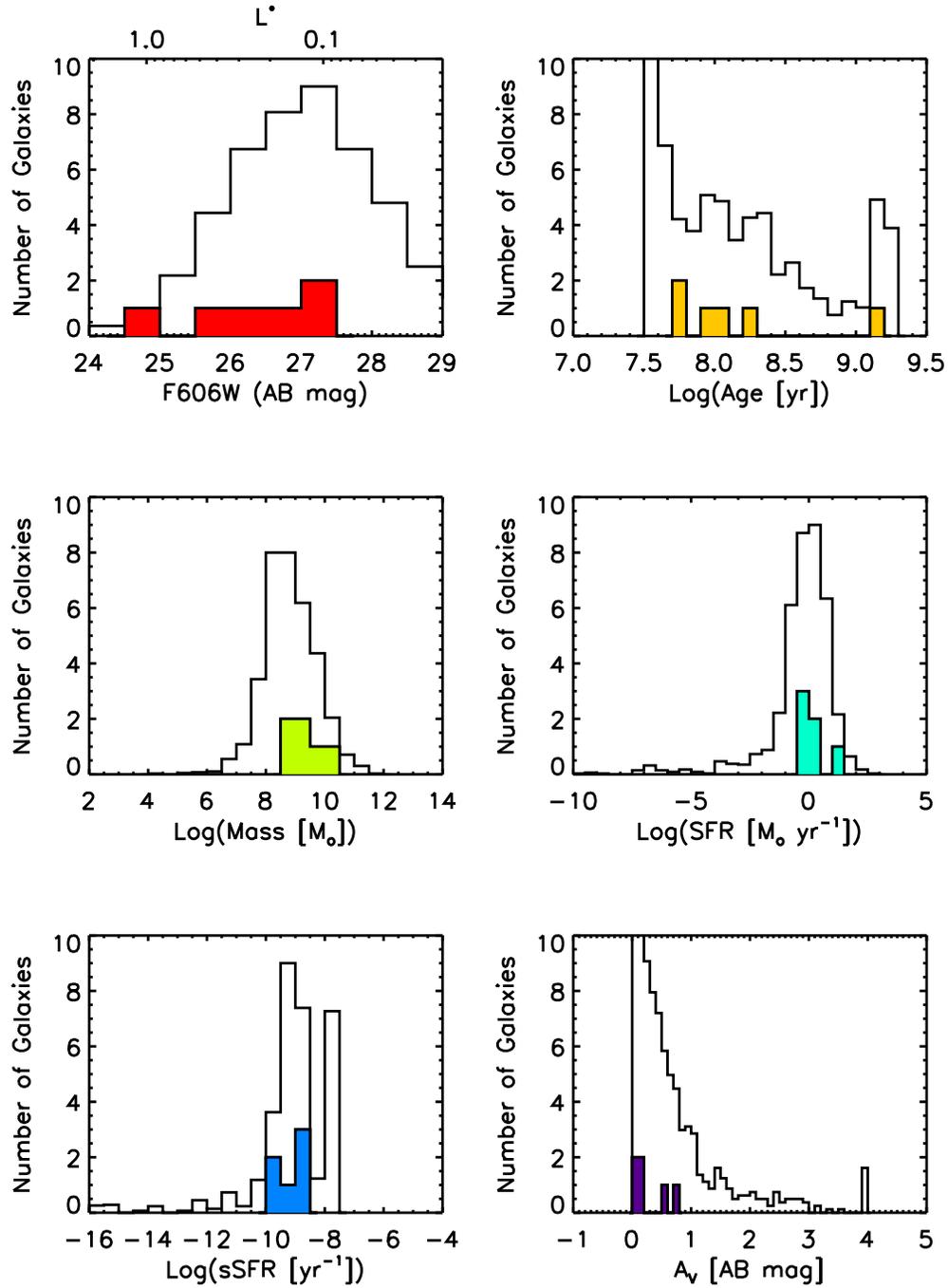} 
\caption[]{
Luminosity function and physical properties derived from SED fitting \citep{skelton14} for galaxies associated 
with the \lya\ nebula (filled histograms) versus all galaxies in the field 
within $\Delta z \pm0.15$ (open histograms, scaled down for clarity). 
{\bf Top Left:} $F606W$-band luminosity function (AB mag). 
{\bf Top Right:} Log of the stellar age in years.
{\bf Middle Left:} Log of the stellar mass in solar masses. 
{\bf Middle Right:} Log of the star formation rate in solar masses per year. 
{\bf Bottom Left:} Log of the specific star formation rate in inverse years.
{\bf Bottom Right:} Attenuation $A_{V}$ (AB mag).
}
\label{fig:galproperties}
\end{figure}

\begin{figure}[th]
\center
\includegraphics[angle=0,width=4.5in]{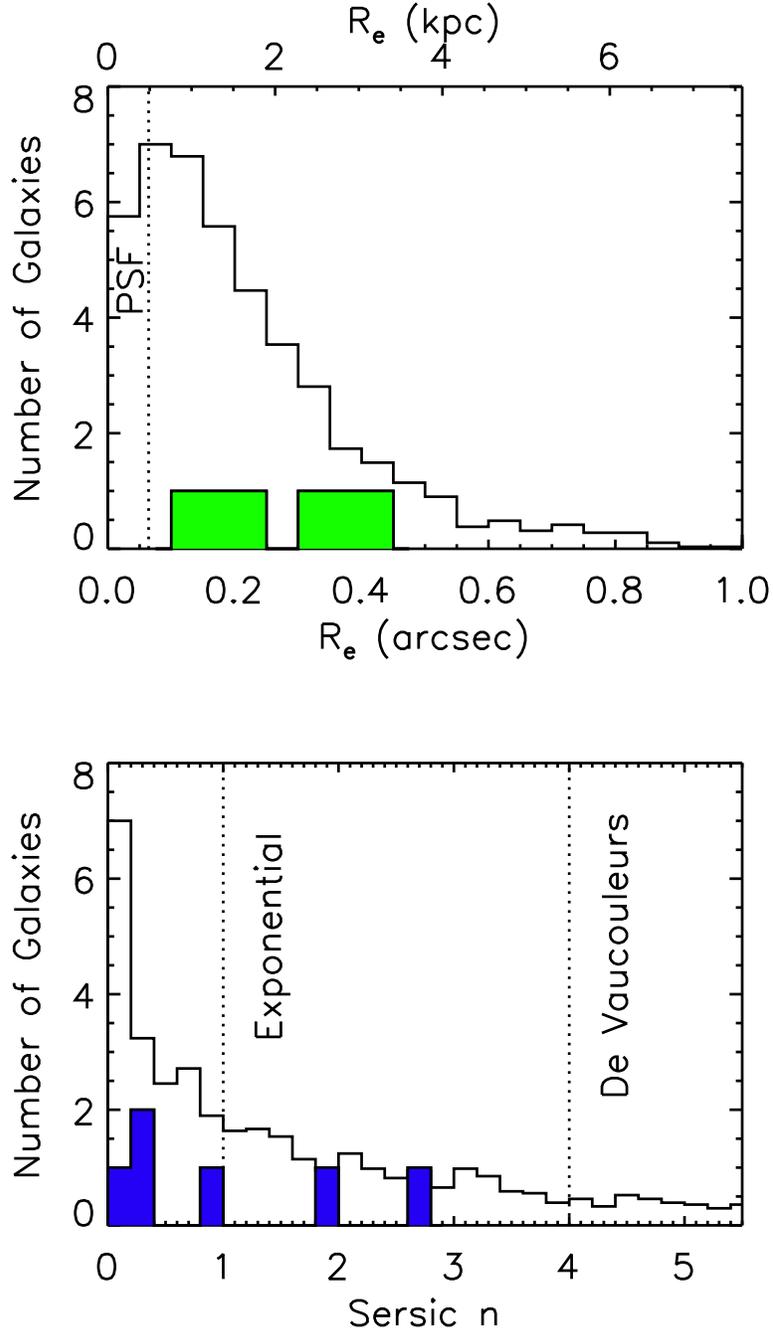} 
\caption[]{
Effective radius ($R_{e}$) and S\'ersic $n$ estimates derived from 
GALFIT parametric fits to the $F160W$ data \citep{vanderwel2012}, for 
member galaxies ($\Delta z\pm0.15, R<\radius\arcsec$; filled histogram), 
and for all galaxies in the GOODS-S field within $\Delta z\pm0.15$ of the nebula redshift (open histogram, 
scaled down for clarity).  The size of the PSF (top panel) and the S\'ersic $n$ values for 
exponential and De Vaucouleurs profiles (bottom panel) are indicated with dotted lines. 
}
\label{fig:galfitmorph}
\end{figure}

Having shown that Source 6 is an obscured AGN likely associated with the nebula, we briefly turn our 
attention to the properties of the other galaxies in this overdense region, and whether they 
are distinctive in any way relative to the field population.   
In Figure~\ref{fig:membersedall} we show the measured photometry and best-fit SEDs 
of the \numgal\ galaxies that are likely associated with the \lya\ nebula. 
We note that none of these member galaxies 
is detected in the MIPS imaging \citep{whitaker2014}.  
Figure~\ref{fig:galproperties} shows the approximate luminosity function (LF) 
for the system, assuming that all \numgal\ galaxies are at $z\approx3.157$, as well as 
the ages, stellar masses, star formation rates, specific star formation rates, and 
$A_{V}$ estimates from stellar population synthesis model fitting using FAST \citep{kriek09}, 
as reported in the 3D-HST/CANDELS catalog \citep{skelton14}.  
The galaxies are typically low luminosity, around 
$\sim0.2$ L$^{*}$, with only one (Source 3) around L$^{*}$, 
assuming that M$^{*}\approx-20.8$ AB mag at this redshift \citep{red08}.  
The total UV luminosity of the member galaxies is 
$\approx$23.9 mag (AB) in the $F606W$ band, or $\approx$2.3 L$^{*}$, 
suggesting this may be a small group in formation.  
The member galaxies have typical ages of around $10^{8}$ yr, stellar masses of 
$10^{9}$ M$_{\odot}$, star formation rates of $<$10 M$_{\odot}$ yr$^{-1}$, 
specific star formation rates of $10^{-9}$ yr$^{-1}$, and extinctions of 
$A_{V}\approx0.2-0.8$ mag.  
In Figure~\ref{fig:galfitmorph}, we show the effective radii ($R_{e}$) and S\'ersic parameter 
($n$) measurements from parametric fits to the $F160W$ imaging for the galaxies in the 
neighborhood of the nebula \citep{vanderwel2012}.  
The member galaxies are typically 1-3~kpc in radius, with S\'ersic parameters clustered more around 
$n\approx1$, the value for an exponential disk, than around $n\approx4$, the value for a De Vaucouleurs profile.  
Two-sample Kolmogorov-Smirnov (KS) tests on all parameters did not reveal any statistical differences between the member galaxies and 
the field population at the same redshift 
($P_{LF}=0.17$, $P_{log(Age)}=0.37$, $P_{log(Mass)}=0.12$, $P_{log(SFR)}=0.34$, $P_{log(sSFR)}=0.92$, $P_{A_{V}}=0.57$, $P_{R_{e}}=0.54$, $P_{n}=0.53$).  
This may be due to the small sample size, or it could reflect that we are witnessing 
an early stage in the formation of a group-mass system when the individual member galaxies are separated by 
large physical distances and environmental effects have yet to become important. 

\section{Powering the \lya\ Nebula}
\label{sec:powering}

\subsection{The Energetics: Revisited}
\label{sec:energetics}

\citet{nil06} estimated that even if Source 3, the brightest neighbor galaxy, was at the nebula 
redshift, it was unlikely to be able to power the \lya\ emission unless highly collimated in the direction 
of the nebula.  
Extrapolating the HST $B$ and $V$-band measurements of all the member galaxies, and accounting for 
the angle subtended by the nebula in each case, we find that galaxies can account for at most 
$\approx$14\% 
of the \lya\ emission, even under the optimistic assumption of an escape fraction of unity. 
Thus, the only source associated with the nebula that could plausibly contribute enough ionizing 
photons is the obscured AGN, i.e., Source 6.

The Spitzer SED of Source 6 (Figure~\ref{fig:source6}) is consistent with a power-law spectrum of: 
$L_{\nu}\approx1.7\times10^{30} (\nu/10^{14})^{-1.75}$ erg s$^{-1}$ Hz$^{-1}$; this corresponds to a 
total infrared luminosity of around $\sim1.5\times10^{45}$ erg s$^{-1}$, estimated by scaling the template 
of Mrk~231 to match the 8$\mu$m flux, which is approaching the realm of ultra-luminous infrared galaxies (ULIRGs; $10^{12} L_{\odot}\approx4\times10^{45}$ erg s$^{-1}$).  
Making the crude assumption that the MIR power-law can be extrapolated into the restframe UV, we 
estimate the ionizing photon flux of Source 6 to be $\sim3.5\times10^{53}$ photons s$^{-1}$ 
between restframe 200 and 912\AA.  
Producing the observed \lya\ luminosity of the nebula ($1.09\pm0.07\times10^{43}$ erg s$^{-1}$) requires 
a total of $\sim1.0\times10^{54}$ photons s$^{-1}$, assuming that two-thirds of recombinations result 
in a \lya\ photon.  Naively, it would appear that Source 6 can only account for about 35\% of the \lya\ emission, 
even when ignoring geometrical dilution.  
However, this extrapolation is very uncertain; for example, a somewhat shallower power-law index of -1.5 is 
sufficient to match the required ionizing photon flux.  If we instead take the total infrared 
luminosity (i.e., 3-1000$\mu$m) to be measure of the intrinsic UV/optical emission 
(i.e., 1000\AA-1$\mu$m) that is being reprocessed by dust, we estimate 
that the intrisic ionizing photon flux from Source 6 is around $\sim10^{55}$ 
photons s$^{-1}$, more than sufficient to power the nebula.
Given the uncertainties about the geometry of the system, the opening angle of the AGN, the obscuration of 
the source, and the true SED shortward of the Lyman limit, it seems plausible that Source 6 
could simply be highly obscured along our line-of-sight but less obscured in the directions 
of the diffuse \lya\ emission.  Therefore, combined with the morphological evidence in 
Section~\ref{sec:spatial}, it appears likely that Source 6 is responsible for a 
significant fraction of the ionization in the \lya\ nebula. 

While we have shown that there are several associated galaxies and an obscured AGN in this system 
that could contribute to the ionization of the nebula, we have also confirmed the previous result that 
there is no compact continuum source visible within the brightest part of the \lya\ nebula itself.  
The GOODS-S HST imaging provides an upper limit on the amount of star formation that could be going on within the nebula: 
the 3$\sigma$ limit on the $F606W$ flux (restframe $\sim$1400\AA) of $4.66\times10^{-30}$ that was 
quoted in the original paper corresponds to a limit on the UV star formation rate of 
$SFR(UV)=1.4\times10^{-28}L_{\nu}<57$ M$_{\odot}$ yr$^{-1}$, assuming the standard 
conversion \citep{ken98}. 
High resolution millimeter/submillimeter imaging will be important for putting further constraints on 
the obscured star formation and molecular gas reservoir within the nebula.  

\subsection{Is the \lya\ nebula gravitationally powered?}

The argument up until now has been that a \lya\ nebula without any associated continuum counterparts is the 
best candidate for a system powered by gravitational cooling radiation.  In view of our improved observational 
understanding of the \citet{nil06} \lya\ nebula 
as well as updated theoretical simulations, it appears that this line of reasoning 
should actually be reversed, and that even this ``best candidate'' for a gravitationally powered \lya\ nebula is in fact 
powered by an AGN.

A number of authors have investigated whether gravitational cooling radition within cold accretion streams can 
explain the observed properties of \lya\ nebulae, initially with analytic arguments \citep{hai00} and then using 
numerical simulations \citep{far01,fur05,yang06,dijk06a,dijk09,goerdt10,faucher10,rosdahl12}.  
Due to the difficulty in appropriately simulating the self-shielding of the gas and the temperature of the IGM, 
the maximum predicted luminosities of ``cooling'' \lya\ nebulae from different numerical treatments have ranged 
from $\sim$10$^{42}$ erg s$^{-1}$, far below that of observed samples, to 
$\sim$10$^{44}$ erg s$^{-1}$, sufficient to explain even the brightest known systems.  
Thus it remains a matter of some debate whether gravitational cooling alone 
can in fact explain the large \lya\ luminosities of \lya\ nebulae.  

At the same time, it is becoming clear that because gravitationally powered \lya\ emission 
originates from the extended cold streams that are fueling galaxy formation, 
star formation, and AGN activity, the simulations generically predict that there will be a growing galaxy 
at the center of any luminous ``cooling" \lya\ nebula 
\citep[e.g.,][]{far01,fur05,goerdt10,faucher10,rosdahl12}.  
Furthermore, these luminous \lya\ nebulae will necessarily be hosted by $\gtrsim10^{12}$ M$_{\odot}$ halos \citep{rosdahl12}.  
While it is possible that such halos go through an initial phase of gas cooling preceding any substantial 
star formation, this cooling-only phase would necessarily have occurred much earlier, when the halo was 
$\sim10^{9}-10^{10}$ M$_{\odot}$ and when the corresponding cooling \lya\ emission was several orders of 
magnitude fainter than observed \lya\ nebulae \citep{fur05}. 
According to the most recent simulations, stellar continuum emission would therefore be expected 
at the position of the \lya\ peak, unless this galactic counterpart has been 
very effectively hidden \citep{rosdahl12}. 
For this reason, we suspect that the correct argument is the opposite of what has often been assumed: 
namely, that the lack of any associated continuum sources in the \lya\ nebula makes it 
{\it less likely} that this system is predominantly powered by gravitational cooling.  
It is difficult to understand how the central growing galaxy, presumably the most massive member in the 
region located at a node within the cosmic web, has been so effectively hidden at the center of the \lya\ 
nebula.  If it is simply highly obscured and therefore not visible in the restframe optical, then it 
should be the brightest source visible in the system at longer wavelengths.  
Higher resolution millimeter/submillimeter observations will be important for shedding further 
light on this issue, but the evidence thus far from the Spitzer data is that the bulk of the 
obscured emission in this system is concentrated at the position of Source 6, an obscured AGN 
that is offset by $\sim4\arcsec\approx30$~kpc from the brightest part of the \lya\ nebula 
and that is instead located in an apparent ``hole'' in the fainter, larger scale \lya\ emission. 
Thus, while it is still possible that the \lya\ emission is tracing gas within cosmic filaments, 
the results of this paper argue strongly that it 
is being powered primarily by illumination from the obscured AGN in the region instead of by gravitational cooling.

We suspect that this \lya\ nebula is not an isolated case.  
The key observational characteristics of this system - 
the small sizes, disk-like morphologies, and low luminosities of the member 
galaxies, the evidence for an overdense neighborhood, and the presence of a 
highly obscured AGN offset by 10s of kpc from the \lya\ peak that can provide 
some if not all of the requisite ionizing photons - are highly reminiscent of 
the properties LABd05, a more luminous \lya\ nebula at $z\approx2.7$ \citep{pres12b}.  
It seems likely that these systems are intrinsically similar to cases such as \citet{wei04,wei05}, a quasar 
at $z\approx3$ with an extended, asymmetric biconical \lya\ halo that is being viewed through one side of 
the ionization cone of the (unobscured) quasar.  In the \citet{nil06} \lya\ nebula, we are likely observing a similar phenomenon, 
an AGN powering an extended \lya\ nebula, but with a line-of-sight passing outside the ionization cone.  
This picture is broadly consistent with the much higher obscuration of the AGN and 
the two lobe structure seen at low surface brightness levels in the \lya\ emission.  
Our conclusions are consistent with the argument of \citet{overzier13} that AGN must be 
the dominant power source for the majority (if not all) of the most luminous \lya\ nebulae.  
The \citet{nil06} Lya nebula is somewhat lower luminosity than the systems studied by \citet{overzier13}, 
suggesting that even at lower luminosities where less powerful mechanisms (gravitational cooling, winds, 
star formation) could in principle play a larger role, AGN continue to be a dominant power source for \lya\ nebulae.

\section{Implications}
\label{sec:implications}

On a more general level, while gravitational cooling radiation, as a fundamental physical process, must 
contribute to the total \lya\ emission observed from \lya\ nebulae, 
our experience with this \lya\ nebula leads us to conclude that it will always be more difficult to prove 
that a given nebula is primarily gravitationally powered.  
Since bright \lya\ emission from gravitationally cooling correlates strongly with actively growing 
galaxies (and AGN), a lack of obvious power sources is not sufficient evidence, and the morphology and 
shape of the surface brightness profile could simply reflect the fact that the gas is being brought in 
by a cold accretion stream without necessarily proving the nebula is lit up by gravitational cooling.  
Determining what fraction of a given \lya\ nebula is powered by gravity as 
opposed to by embedded sources requires more detailed studies of, e.g., the 
polarization of the \lya\ emission to constrain the contribution of scattered light, as well as 
comparisons between emission line diagnostic measurements within the 
diffuse gas and improved theoretical predictions for the expected 
metal line emission from cold accrection streams. 

Instead, we suspect that extended \lya\ emission displaced from all associated continuum 
sources is actually {\it less likely} to be powered by cooling radition and {\it more likely} to be a signature 
of cosmic gas illuminated by an offset AGN (leading to either \lya\ fluorescence or \lya\ resonant scattering), 
an AGN that may be highly-obscured to the line-of-sight or recently extinguished.  
It is not particularly hard to hide an AGN with obscuration and geometry (i.e., if it is 
offset from the nebula and obscured to our line-of-sight but unobscured towards the nebula), or 
if it is in the process of ramping down in its output, leaving a seemingly-abandoned 
light echo of \lya\ emission at larger distances (and light travel times) from the 
AGN position \citep[e.g., similar to Hanny's Voorwerp, an \oiii\ nebula 
at lower redshift;][]{schawinski2010}.  
In systems where we can identify an AGN as the dominant power source in this way, we have an opportunity to 
use the spatially extended line emission as a laboratory for studying AGN variability over 
$10^4$-$10^5$ year timescales as well as the physical conditions within the underlying cosmic web, 
lit up in emission while in the process of fueling and being impacted by ongoing local galaxy formation.


\section{Conclusions}
\label{sec:conclusions}

In light of new observations and more sophisticated numerical simulations, we have taken a second look 
at the local environment and most likely power source for the \citet{nil06} \lya\ nebula.  
We find that the nebula is associated with \numgal\ nearby galaxies and an obscured AGN located 
$\sim$4\arcsec$\approx$30~kpc away.  The \lya\ emission is seen to extend further than previously realized, 
nearly encircling the position of the obscured AGN, and the local region shows evidence for being 
overdense relative to the field.  At the same time, results from the latest generation 
of cold accretion simulations show that the brightest \lya\ emission powered by gravitational cooling will be centered 
on a growing galaxy, which should be detectable as a counterpart in the continuum.  
Thus, while we confirm the previous finding that there is no associated continuum source located within the 
nebula itself, we conclude that this is actually a strong argument {\it against}, rather than for, the nebula being 
powered by gravitational energy.  The \lya\ emission at low surface brightness levels appears to trace out 
a complicated filamentary structure, perhaps consistent with gas within cold accretion streams, but this gas is 
most plausibly being lit up due to illumination from a nearby obscured AGN rather than due to gravitational powering.  
Follow-up observations of such emission line nebulae offer an exciting window into both AGN physics and 
the kinematics and physical conditions of the cosmic web surrounding regions of active galaxy formation.

\acknowledgments
We are grateful to Kim Nilsson for assistance with the original \lya\ data, 
to Rosalind Skelton for guidance on using the 3D-HST/CANDELS photometric catalog, 
to Julie Wardlow and the HerMES team for providing early access to the HerMES catalogs, 
and to Lise Christensen, Kristian Finlator, Sebastian H\"onig, Pieter van Dokkum, 
Natascha F{\"o}rster Schreiber, and the anonymous referee for helpful discussions and comments.

M.K.M.P. was supported by a Dark Cosmology Centre Fellowship.  J.P.U.F. acknowledges support 
from the ERC-StG grant EGGS-278202.  The Dark Cosmology Centre is funded by The Danish National 
Research Foundation.  

This work is based on observations taken by 
the WFC3 Early Release Science program (GO 11359 and 11360) 
and the 3D-HST Treasury Program (GO 12177 and 12328) 
with the NASA/ESA HST, which is operated by the Association of Universities for Research in Astronomy, Inc., 
under NASA contract NAS5-26555.

The GOODS-Herschel data was accessed through the Herschel Database in Marseille 
(HeDaM - http://hedam.lam.fr) operated by CeSAM and hosted by the Laboratoire d'Astrophysique de Marseille.

This research has made use of data from HerMES project (http://hermes.sussex.ac.uk/).  
HerMES is a Herschel Key Programme utilising Guaranteed Time from the SPIRE instrument team, 
ESAC scientists and a mission scientist.  


\begin{thebibliography}{72}
\expandafter\ifx\csname natexlab\endcsname\relax\def\natexlab#1{#1}\fi

\bibitem[{Adams {et~al.}(2009)Adams, Hill, \& MacQueen}]{adams09}
Adams, J.~J., Hill, G.~J., \& MacQueen, P.~J. 2009, The Astrophysical Journal,
  694, 314

\bibitem[{Brammer {et~al.}(2008)Brammer, van Dokkum, \& Coppi}]{brammer08}
Brammer, G.~B., van Dokkum, P.~G., \& Coppi, P. 2008, The Astrophysical
  Journal, 686, 1503

\bibitem[{Cantalupo {et~al.}(2005)Cantalupo, Porciani, Lilly, \&
  Miniati}]{cantalupo05}
Cantalupo, S., Porciani, C., Lilly, S.~J., \& Miniati, F. 2005, The
  Astrophysical Journal, 628, 61

\bibitem[{Cen \& Zheng(2013)}]{cen13}
Cen, R., \& Zheng, Z. 2013, The Astrophysical Journal, 775, 112

\bibitem[{Colbert {et~al.}(2011)Colbert, Scarlata, Teplitz, Francis, Palunas,
  Williger, \& Woodgate}]{colbert11}
Colbert, J.~W., Scarlata, C., Teplitz, H., Francis, P., Palunas, P., Williger,
  G.~M., \& Woodgate, B. 2011, The Astrophysical Journal, 728, 59

\bibitem[{Dahlen {et~al.}(2013)Dahlen, Mobasher, Faber, Ferguson, Barro,
  Finkelstein, Finlator, Fontana, Gruetzbauch, Johnson, Pforr, Salvato,
  Wiklind, Wuyts, Acquaviva, Dickinson, Guo, Huang, Huang, Newman, Bell,
  Conselice, Galametz, Gawiser, Giavalisco, Grogin, Hathi, Kocevski, Koekemoer,
  Koo, Lee, McGrath, Papovich, Peth, Ryan, Somerville, Weiner, \&
  Wilson}]{dahlen13}
Dahlen, T., {et~al.} 2013, The Astrophysical Journal, 775, 93

\bibitem[{Dekel {et~al.}(2009)Dekel, Birnboim, Engel, Freundlich, Goerdt,
  Mumcuoglu, Neistein, Pichon, Teyssier, \& Zinger}]{dekel09}
Dekel, A., {et~al.} 2009, Nature, 457, 451

\bibitem[{Dickinson {et~al.}(2004)Dickinson, Stern, Giavalisco, Ferguson,
  Tsvetanov, Chornock, Cristiani, Dawson, Dey, Filippenko, Moustakas, Nonino,
  Papovich, Ravindranath, Riess, Rosati, Spinrad, \& Vanzella}]{dickinson2004}
Dickinson, M., {et~al.} 2004, The Astrophysical Journal, 600, L99

\bibitem[{Dijkstra(2009)}]{dijk09}
Dijkstra, M. 2009, The Astrophysical Journal, 690, 82

\bibitem[{Dijkstra {et~al.}(2006)Dijkstra, Haiman, \& Spaans}]{dijk06a}
Dijkstra, M., Haiman, Z., \& Spaans, M. 2006, The Astrophysical Journal, 649,
  14

\bibitem[{Donley {et~al.}(2008)Donley, Rieke, P\'{e}rez-Gonz\'{a}lez, \&
  Barro}]{don08}
Donley, J.~L., Rieke, G.~H., P\'{e}rez-Gonz\'{a}lez, P.~G., \& Barro, G. 2008,
  The Astrophysical Journal, 687, 111

\bibitem[{Donley {et~al.}(2007)Donley, Rieke, Perez‐Gonzalez, Rigby, \&
  Alonso‐Herrero}]{donley07}
Donley, J.~L., Rieke, G.~H., Perez‐Gonzalez, P.~G., Rigby, J.~R., \&
  Alonso‐Herrero, A. 2007, The Astrophysical Journal, 660, 167

\bibitem[{Donley {et~al.}(2012)Donley, Koekemoer, Brusa, Capak, Cardamone,
  Civano, Ilbert, Impey, Kartaltepe, Miyaji, Salvato, Sanders, Trump, \&
  Zamorani}]{don12}
Donley, J.~L., {et~al.} 2012, The Astrophysical Journal, 748, 142

\bibitem[{Elbaz {et~al.}(2011)Elbaz, Dickinson, Hwang, D\'{\i}az-Santos,
  Magdis, Magnelli, {Le Borgne}, Galliano, Pannella, Chanial, Armus,
  Charmandaris, Daddi, Aussel, Popesso, Kartaltepe, Altieri, Valtchanov, Coia,
  Dannerbauer, Dasyra, Leiton, Mazzarella, Alexander, Buat, Burgarella, Chary,
  Gilli, Ivison, Juneau, {Le Floc’h}, Lutz, Morrison, Mullaney, Murphy, Pope,
  Scott, Brodwin, Calzetti, Cesarsky, Charlot, Dole, Eisenhardt, Ferguson,
  {F\"{o}rster Schreiber}, Frayer, Giavalisco, Huynh, Koekemoer, Papovich,
  Reddy, Surace, Teplitz, Yun, \& Wilson}]{elbaz11}
Elbaz, D., {et~al.} 2011, Astronomy \& Astrophysics, 533, A119

\bibitem[{Erb {et~al.}(2011)Erb, Bogosavljevi\'{c}, \& Steidel}]{erb11}
Erb, D.~K., Bogosavljevi\'{c}, M., \& Steidel, C.~C. 2011, The Astrophysical
  Journal, 740, L31

\bibitem[{Fardal {et~al.}(2001)Fardal, Katz, Gardner, Hernquist, Weinberg, \&
  Dave}]{far01}
Fardal, M.~A., Katz, N., Gardner, J.~P., Hernquist, L., Weinberg, D.~H., \&
  Dave, R. 2001, The Astrophysical Journal, 562, 605

\bibitem[{Faucher-Gigu\`{e}re {et~al.}(2010)Faucher-Gigu\`{e}re, Kere\v{s},
  Dijkstra, Hernquist, \& Zaldarriaga}]{faucher10}
Faucher-Gigu\`{e}re, C.-A., Kere\v{s}, D., Dijkstra, M., Hernquist, L., \&
  Zaldarriaga, M. 2010, The Astrophysical Journal, 725, 633

\bibitem[{Furlanetto {et~al.}(2005)Furlanetto, Schaye, Springel, \&
  Hernquist}]{fur05}
Furlanetto, S.~R., Schaye, J., Springel, V., \& Hernquist, L. 2005, The
  Astrophysical Journal, 622, 7

\bibitem[{Geach {et~al.}(2009)Geach, Alexander, Lehmer, Smail, Matsuda,
  Chapman, Scharf, Ivison, Volonteri, Yamada, Blain, Bower, Bauer, \&
  Basu-Zych}]{gea09}
Geach, J.~E., {et~al.} 2009, The Astrophysical Journal, 700, 1

\bibitem[{Giavalisco {et~al.}(2004)Giavalisco, Ferguson, Koekemoer, Dickinson,
  Alexander, Bauer, Bergeron, Biagetti, Brandt, Casertano, Cesarsky,
  Chatzichristou, Conselice, Cristiani, {Da Costa}, Dahlen, de~Mello,
  Eisenhardt, Erben, Fall, Fassnacht, Fosbury, Fruchter, Gardner, Grogin, Hook,
  Hornschemeier, Idzi, Jogee, Kretchmer, Laidler, Lee, Livio, Lucas, Madau,
  Mobasher, Moustakas, Nonino, Padovani, Papovich, Park, Ravindranath, Renzini,
  Richardson, Riess, Rosati, Schirmer, Schreier, Somerville, Spinrad, Stern,
  Stiavelli, Strolger, Urry, Vandame, Williams, \& Wolf}]{giav04}
Giavalisco, M., {et~al.} 2004, The Astrophysical Journal, 600, L93

\bibitem[{Goerdt {et~al.}(2010)Goerdt, Dekel, Sternberg, Ceverino, Teyssier, \&
  Primack}]{goerdt10}
Goerdt, T., Dekel, A., Sternberg, A., Ceverino, D., Teyssier, R., \& Primack,
  J.~R. 2010, Monthly Notices of the Royal Astronomical Society, 407, 613

\bibitem[{Haiman {et~al.}(2000)Haiman, Spaans, \& Quataert}]{hai00}
Haiman, Z., Spaans, M., \& Quataert, E. 2000, The Astrophysical Journal, 537,
  L5

\bibitem[{Hayes {et~al.}(2011)Hayes, Scarlata, \& Siana}]{hayes11}
Hayes, M., Scarlata, C., \& Siana, B. 2011, Nature, 476, 304

\bibitem[{Ivison {et~al.}(2004)Ivison, Greve, Serjeant, Bertoldi, Egami,
  Mortier, Alonso‐Herrero, Barmby, Bei, Dole, Engelbracht, Fazio, Frayer,
  Gordon, Hines, Huang, {Le Floc’h}, Misselt, Miyazaki, Morrison, Papovich,
  Perez‐Gonzalez, Rieke, Rieke, Rigby, Rigopoulou, Smail, Wilson, \&
  Willner}]{ivison04}
Ivison, R.~J., {et~al.} 2004, The Astrophysical Journal Supplement Series, 154,
  124

\bibitem[{Kennicutt(1998)}]{ken98}
Kennicutt, R.~C. 1998, The Astrophysical Journal, 498, 541

\bibitem[{Keres {et~al.}(2005)Keres, Katz, Weinberg, \& Dave}]{keres05}
Keres, D., Katz, N., Weinberg, D.~H., \& Dave, R. 2005, Monthly Notices of the
  Royal Astronomical Society, 363, 2

\bibitem[{Kollmeier {et~al.}(2010)Kollmeier, Zheng, Dav\'{e}, Gould, Katz,
  Miralda-Escud\'{e}, \& Weinberg}]{koll10}
Kollmeier, J.~A., Zheng, Z., Dav\'{e}, R., Gould, A., Katz, N.,
  Miralda-Escud\'{e}, J., \& Weinberg, D.~H. 2010, The Astrophysical Journal,
  708, 1048

\bibitem[{Kriek {et~al.}(2009)Kriek, van Dokkum, Labb\'{e}, Franx, Illingworth,
  Marchesini, \& Quadri}]{kriek09}
Kriek, M., van Dokkum, P.~G., Labb\'{e}, I., Franx, M., Illingworth, G.~D.,
  Marchesini, D., \& Quadri, R.~F. 2009, The Astrophysical Journal, 700, 221

\bibitem[{Lacy {et~al.}(2007)Lacy, Petric, Sajina, Canalizo, Storrie-Lombardi,
  Armus, Fadda, \& Marleau}]{lac07}
Lacy, M., Petric, A.~O., Sajina, A., Canalizo, G., Storrie-Lombardi, L.~J.,
  Armus, L., Fadda, D., \& Marleau, F.~R. 2007, The Astronomical Journal, 133,
  186

\bibitem[{Lacy {et~al.}(2004)Lacy, Storrie-Lombardi, Sajina, Appleton, Armus,
  Chapman, Choi, Fadda, Fang, Frayer, Heinrichsen, Helou, Im, Marleau, Masci,
  Shupe, Soifer, Surace, Teplitz, Wilson, \& Yan}]{lac04}
Lacy, M., {et~al.} 2004, The Astrophysical Journal Supplement Series, 154, 166

\bibitem[{Laursen {et~al.}(2013)Laursen, Duval, \& \"{O}stlin}]{laursen2013}
Laursen, P., Duval, F., \& \"{O}stlin, G. 2013, The Astrophysical Journal, 766,
  124

\bibitem[{Laursen \& Sommer-Larsen(2007)}]{laursen07}
Laursen, P., \& Sommer-Larsen, J. 2007, The Astrophysical Journal, 657, L69

\bibitem[{Laursen {et~al.}(2009)Laursen, Sommer-Larsen, \&
  Andersen}]{laursen09}
Laursen, P., Sommer-Larsen, J., \& Andersen, A.~C. 2009, The Astrophysical
  Journal, 704, 1640

\bibitem[{Levenson {et~al.}(2010)Levenson, Marsden, Zemcov, Amblard, Blain,
  Bock, Chapin, Conley, Cooray, Dowell, Ellsworth-Bowers, Franceschini, Glenn,
  Griffin, Halpern, Nguyen, Oliver, Page, Papageorgiou, P\'{e}rez-Fournon,
  Pohlen, Rangwala, Rowan-Robinson, Schulz, Scott, Serra, Shupe, Valiante,
  Vieira, Vigroux, Wiebe, Wright, \& Xu}]{levenson10}
Levenson, L., {et~al.} 2010, Monthly Notices of the Royal Astronomical Society,
  409, 83

\bibitem[{Luo {et~al.}(2008)Luo, Bauer, Brandt, Alexander, Lehmer, Schneider,
  Brusa, Comastri, Fabian, Finoguenov, Gilli, Hasinger, Hornschemeier,
  Koekemoer, Mainieri, Paolillo, Rosati, Shemmer, Silverman, Smail, Steffen, \&
  Vignali}]{luo2008}
Luo, B., {et~al.} 2008, The Astrophysical Journal Supplement Series, 179, 19

\bibitem[{M{\o}ller \& Warren(1998)}]{moller98}
M{\o}ller, P., \& Warren, S.~J. 1998, Monthly Notices of the Royal Astronomical
  Society, 299, 661

\bibitem[{Mori {et~al.}(2004)Mori, Umemura, \& Ferrara}]{mori04}
Mori, M., Umemura, M., \& Ferrara, A. 2004, The Astrophysical Journal, 613, L97

\bibitem[{Nilsson {et~al.}(2006)Nilsson, Fynbo, M{\o}ller, Sommer-Larsen, \&
  Ledoux}]{nil06}
Nilsson, K.~K., Fynbo, J. P.~U., M{\o}ller, P., Sommer-Larsen, J., \& Ledoux, C.
  2006, Astronomy and Astrophysics, 452, L23

\bibitem[{Nilsson {et~al.}(2007)Nilsson, M{\o}ller, M\"{o}ller, Fynbo,
  Michałowski, Watson, Ledoux, Rosati, Pedersen, \& Grove}]{nil07}
Nilsson, K.~K., {et~al.} 2007, Astronomy and Astrophysics, 471, 71

\bibitem[{Oke(1974)}]{oke74}
Oke, J.~B. 1974, The Astrophysical Journal Supplement Series, 27, 21

\bibitem[{Oliver {et~al.}(2012)Oliver, Bock, Altieri, Amblard, Arumugam,
  Aussel, Babbedge, Beelen, B\'{e}thermin, Blain, Boselli, Bridge, Brisbin,
  Buat, Burgarella, Castro-Rodr\'{\i}guez, Cava, Chanial, Cirasuolo, Clements,
  Conley, Conversi, Cooray, Dowell, Dubois, Dwek, Dye, Eales, Elbaz, Farrah,
  Feltre, Ferrero, Fiolet, Fox, Franceschini, Gear, Giovannoli, Glenn, Gong,
  {Gonz\'{a}lez Solares}, Griffin, Halpern, Harwit, Hatziminaoglou, Heinis,
  Hurley, Hwang, Hyde, Ibar, Ilbert, Isaak, Ivison, Lagache, {Le Floc'h},
  Levenson, Faro, Lu, Madden, Maffei, Magdis, Mainetti, Marchetti, Marsden,
  Marshall, Mortier, Nguyen, O'Halloran, Omont, Page, Panuzzo, Papageorgiou,
  Patel, Pearson, P\'{e}rez-Fournon, Pohlen, Rawlings, Raymond, Rigopoulou,
  Riguccini, Rizzo, Rodighiero, Roseboom, Rowan-Robinson, {S\'{a}nchez Portal},
  Schulz, Scott, Seymour, Shupe, Smith, Stevens, Symeonidis, Trichas, Tugwell,
  Vaccari, Valtchanov, Vieira, Viero, Vigroux, Wang, Ward, Wardlow, Wright, Xu,
  \& Zemcov}]{oliver12}
Oliver, S.~J., {et~al.} 2012, Monthly Notices of the Royal Astronomical
  Society, 424, 1614

\bibitem[{Overzier {et~al.}(2013)Overzier, Nesvadba, Dijkstra, Hatch, Lehnert,
  Villar-Mart\'{\i}n, Wilman, \& Zirm}]{overzier13}
Overzier, R.~A., Nesvadba, N. P.~H., Dijkstra, M., Hatch, N.~A., Lehnert,
  M.~D., Villar-Mart\'{\i}n, M., Wilman, R.~J., \& Zirm, A.~W. 2013, The
  Astrophysical Journal, 771, 89

\bibitem[{Polletta {et~al.}(2007)Polletta, Tajer, Maraschi, Trinchieri,
  Lonsdale, Chiappetti, Andreon, Pierre, {Le Fevre}, Zamorani, Maccagni,
  Garcet, Surdej, Franceschini, Alloin, Shupe, Surace, Fang, Rowan‐Robinson,
  Smith, \& Tresse}]{polletta07}
Polletta, M., {et~al.} 2007, The Astrophysical Journal, 663, 81

\bibitem[{Prescott {et~al.}(2009)Prescott, Dey, \& Jannuzi}]{pres09}
Prescott, M. K.~M., Dey, A., \& Jannuzi, B.~T. 2009, The Astrophysical Journal,
  702, 554

\bibitem[{Prescott {et~al.}(2012{\natexlab{a}})Prescott, Dey, \&
  Jannuzi}]{pres12a}
Prescott, M. K.~M., Dey, A., \& Jannuzi, B.~T. 2012{\natexlab{a}}, The
  Astrophysical Journal, 748, 125

\bibitem[{Prescott {et~al.}(2013)Prescott, Dey, \& Jannuzi}]{pres13}
Prescott, M. K.~M., Dey, A., \& Jannuzi, B.~T. 2013, The Astrophysical Journal,
  762, 38

\bibitem[{Prescott {et~al.}(2012{\natexlab{b}})Prescott, Dey, Brodwin, Chaffee,
  Desai, Eisenhardt, {Le Floc'h}, Jannuzi, Kashikawa, Matsuda, \&
  Soifer}]{pres12b}
Prescott, M. K.~M., {et~al.} 2012{\natexlab{b}}, The Astrophysical Journal,
  752, 86

\bibitem[{Rasmussen {et~al.}(2009)Rasmussen, Sommer-Larsen, Pedersen, Toft,
  Benson, Bower, \& Grove}]{rasmussen09}
Rasmussen, J., Sommer-Larsen, J., Pedersen, K., Toft, S., Benson, A., Bower,
  R.~G., \& Grove, L.~F. 2009, The Astrophysical Journal, 697, 79

\bibitem[{Reddy {et~al.}(2008)Reddy, Steidel, Pettini, Adelberger, Shapley,
  Erb, \& Dickinson}]{red08}
Reddy, N.~A., Steidel, C.~C., Pettini, M., Adelberger, K.~L., Shapley, A.~E.,
  Erb, D.~K., \& Dickinson, M. 2008, The Astrophysical Journal Supplement
  Series, 175, 48

\bibitem[{Rosdahl \& Blaizot(2012)}]{rosdahl12}
Rosdahl, J., \& Blaizot, J. 2012, Monthly Notices of the Royal Astronomical
  Society, 423, 344

\bibitem[{Roseboom {et~al.}(2010)Roseboom, Oliver, Kunz, Altieri, Amblard,
  Arumugam, Auld, Aussel, Babbedge, B\'{e}thermin, Blain, Bock, Boselli,
  Brisbin, Buat, Burgarella, Castro-Rodr\'{\i}guez, Cava, Chanial, Chapin,
  Clements, Conley, Conversi, Cooray, Dowell, Dwek, Dye, Eales, Elbaz, Farrah,
  Fox, Franceschini, Gear, Glenn, Solares, Griffin, Halpern, Harwit,
  Hatziminaoglou, Huang, Ibar, Isaak, Ivison, Lagache, Levenson, Lu, Madden,
  Maffei, Mainetti, Marchetti, Marsden, Mortier, Nguyen, O'Halloran, Omont,
  Page, Panuzzo, Papageorgiou, Patel, Pearson, P\'{e}rez-Fournon, Pohlen,
  Rawlings, Raymond, Rigopoulou, Rizzo, Rowan-Robinson, Portal, Schulz, Scott,
  Seymour, Shupe, Smith, Stevens, Symeonidis, Trichas, Tugwell, Vaccari,
  Valtchanov, Vieira, Vigroux, Wang, Ward, Wright, Xu, \&
  Zemcov}]{roseboom2010}
Roseboom, I.~G., {et~al.} 2010, Monthly Notices of the Royal Astronomical
  Society, 409, 48

\bibitem[{Roseboom {et~al.}(2012)Roseboom, Ivison, Greve, Amblard, Arumugam,
  Auld, Aussel, Bethermin, Blain, Bock, Boselli, Brisbin, Buat, Burgarella,
  Castro-Rodr\'{\i}guez, Cava, Chanial, Chapin, Chapman, Clements, Conley,
  Conversi, Cooray, Dowell, Dunlop, Dwek, Eales, Elbaz, Farrah, Franceschini,
  Glenn, Griffin, Halpern, Hatziminaoglou, Ibar, Isaak, Lagache, Levenson, Lu,
  Madden, Maffei, Mainetti, Marchetti, Marsden, Morrison, Mortier, Nguyen,
  O’Halloran, Oliver, Omont, Page, Panuzzo, Papageorgiou, Pearson,
  P\'{e}rez-Fournon, Pohlen, Rawlings, Raymond, Rigopoulou, Rizzo, Rodighiero,
  Rowan-Robinson, Schulz, Scott, Seymour, Shupe, Smith, Stevens, Symeonidis,
  Trichas, Tugwell, Vaccari, Valtchanov, Vieira, Viero, Vigroux, Wardlow, Wang,
  Wright, Xu, \& Zemcov}]{roseboom2012}
Roseboom, I.~G., {et~al.} 2012, Monthly Notices of the Royal Astronomical
  Society, 419, 2758

\bibitem[{Schawinski {et~al.}(2010)Schawinski, Evans, Virani, {Megan Urry},
  Keel, Natarajan, Lintott, Manning, Coppi, Kaviraj, Bamford, J\'{o}zsa,
  Garrett, van Arkel, Gay, \& Fortson}]{schawinski2010}
Schawinski, K., {et~al.} 2010, The Astrophysical Journal, 724, L30

\bibitem[{Skelton {et~al.}(2014)Skelton, Whitaker, Momcheva, Brammer, van
  Dokkum, Labb\'{e}, Franx, van~der Wel, Bezanson, {Da Cunha}, Fumagalli,
  {F\"{o}rster Schreiber}, Kriek, Leja, Lundgren, Magee, Marchesini, Maseda,
  Nelson, Oesch, Pacifici, Patel, Price, Rix, Tal, Wake, \& Wuyts}]{skelton14}
Skelton, R.~E., {et~al.} 2014, The Astrophysical Journal Supplement Series,
  214, 24

\bibitem[{Smith {et~al.}(2008)Smith, Jarvis, Lacy, \&
  Mart\'{\i}nez-Sansigre}]{smi08}
Smith, D. J.~B., Jarvis, M.~J., Lacy, M., \& Mart\'{\i}nez-Sansigre, A. 2008,
  Monthly Notices of the Royal Astronomical Society, 389, 799

\bibitem[{Steidel {et~al.}(2011)Steidel, Bogosavljevi\'{c}, Shapley, Kollmeier,
  Reddy, Erb, \& Pettini}]{stei11}
Steidel, C.~C., Bogosavljevi\'{c}, M., Shapley, A.~E., Kollmeier, J.~A., Reddy,
  N.~A., Erb, D.~K., \& Pettini, M. 2011, The Astrophysical Journal, 736, 160

\bibitem[{Stern {et~al.}(2005)Stern, Eisenhardt, Gorjian, Kochanek, Caldwell,
  Eisenstein, Brodwin, Brown, Cool, Dey, Green, Jannuzi, Murray, Pahre, \&
  Willner}]{stern05}
Stern, D., {et~al.} 2005, The Astrophysical Journal, 631, 163

\bibitem[{Tamura {et~al.}(2013)Tamura, Matsuda, Ikarashi, Scott, Hatsukade,
  Umehata, Saito, Nakanishi, Yun, Ezawa, Hughes, Iono, Kawabe, Kohno, \&
  Wilson}]{tamura2013}
Tamura, Y., {et~al.} 2013, Monthly Notices of the Royal Astronomical Society,
  430, 2768

\bibitem[{Taniguchi \& Shioya(2000)}]{tani00}
Taniguchi, Y., \& Shioya, Y. 2000, The Astrophysical Journal, 532, L13

\bibitem[{Taniguchi {et~al.}(2001)Taniguchi, Shioya, \& Kakazu}]{tani01}
Taniguchi, Y., Shioya, Y., \& Kakazu, Y. 2001, The Astrophysical Journal, 562,
  L15

\bibitem[{van~der Wel {et~al.}(2012)van~der Wel, Bell, H\"{a}ussler, McGrath,
  Chang, Guo, McIntosh, Rix, Barden, Cheung, Faber, Ferguson, Galametz, Grogin,
  Hartley, Kartaltepe, Kocevski, Koekemoer, Lotz, Mozena, Peth, \&
  Peng}]{vanderwel2012}
van~der Wel, A., {et~al.} 2012, The Astrophysical Journal Supplement Series,
  203, 24

\bibitem[{Weidinger {et~al.}(2004)Weidinger, M{\o}ller, \& Fynbo}]{wei04}
Weidinger, M., M{\o}ller, P., \& Fynbo, J. P.~U. 2004, Nature, 430, 999

\bibitem[{Weidinger {et~al.}(2005)Weidinger, M{\o}ller, Fynbo, \&
  Thomsen}]{wei05}
Weidinger, M., M{\o}ller, P., Fynbo, J. P.~U., \& Thomsen, B. 2005, Astronomy
  and Astrophysics, 436, 825

\bibitem[{Weijmans {et~al.}(2010)Weijmans, Bower, Geach, Swinbank, Wilman,
  de~Zeeuw, \& Morris}]{weij10}
Weijmans, A.-M., Bower, R.~G., Geach, J.~E., Swinbank, A.~M., Wilman, R.~J.,
  de~Zeeuw, P.~T., \& Morris, S.~L. 2010, Monthly Notices of the Royal
  Astronomical Society, 402, 2245

\bibitem[{Whitaker {et~al.}(2014)Whitaker, Franx, Leja, van Dokkum, Henry,
  Skelton, Fumagalli, Momcheva, Brammer, Labb\'{e}, Nelson, \&
  Rigby}]{whitaker2014}
Whitaker, K.~E., {et~al.} 2014, The Astrophysical Journal, 795, 104

\bibitem[{Windhorst {et~al.}(2011)Windhorst, Cohen, Hathi, McCarthy, Ryan, Yan,
  Baldry, Driver, Frogel, Hill, Kelvin, Koekemoer, Mechtley, O'Connell,
  Robotham, Rutkowski, Seibert, Straughn, Tuffs, Balick, Bond, Bushouse,
  Calzetti, Crockett, Disney, Dopita, Hall, Holtzman, Kaviraj, Kimble,
  MacKenty, Mutchler, Paresce, Saha, Silk, Trauger, Walker, Whitmore, \&
  Young}]{windhorst2011}
Windhorst, R.~A., {et~al.} 2011, The Astrophysical Journal Supplement Series,
  193, 27

\bibitem[{Xue {et~al.}(2011)Xue, Luo, Brandt, Bauer, Lehmer, Broos, Schneider,
  Alexander, Brusa, Comastri, Fabian, Gilli, Hasinger, Hornschemeier,
  Koekemoer, Liu, Mainieri, Paolillo, Rafferty, Rosati, Shemmer, Silverman,
  Smail, Tozzi, \& Vignali}]{xue2011}
Xue, Y.~Q., {et~al.} 2011, The Astrophysical Journal Supplement Series, 195, 10

\bibitem[{Yajima {et~al.}(2013)Yajima, Li, \& Zhu}]{yajima2013}
Yajima, H., Li, Y., \& Zhu, Q. 2013, The Astrophysical Journal, 773, 151

\bibitem[{Yang {et~al.}(2014)Yang, Walter, Decarli, Bertoldi, Weiss, Dey,
  Prescott, \& Bădescu}]{yang14a}
Yang, Y., Walter, F., Decarli, R., Bertoldi, F., Weiss, A., Dey, A., Prescott,
  M. K.~M., \& Bădescu, T. 2014, The Astrophysical Journal, 784, 171

\bibitem[{Yang {et~al.}(2010)Yang, Zabludoff, Eisenstein, \& Dav\'{e}}]{yang10}
Yang, Y., Zabludoff, A., Eisenstein, D., \& Dav\'{e}, R. 2010, The
  Astrophysical Journal, 719, 1654

\bibitem[{Yang {et~al.}(2006)Yang, Zabludoff, Dave, Eisenstein, Pinto, Katz,
  Weinberg, \& Barton}]{yang06}
Yang, Y., Zabludoff, A.~I., Dave, R., Eisenstein, D.~J., Pinto, P.~A., Katz,
  N., Weinberg, D.~H., \& Barton, E.~J. 2006, The Astrophysical Journal, 640,
  539

\bibitem[{Zheng {et~al.}(2011)Zheng, Cen, Weinberg, Trac, \&
  Miralda-Escud\'{e}}]{zheng11}
Zheng, Z., Cen, R., Weinberg, D., Trac, H., \& Miralda-Escud\'{e}, J. 2011, The
  Astrophysical Journal, 739, 62

\end{thebibliography}

\begin{turnpage}
\begin{deluxetable}{cccccccc}
\tabletypesize{\scriptsize}
\tablecaption{Likely Galaxy Members}
\tablewidth{0pt}
\tablehead{
\colhead{ID\tablenotemark{a}} & \colhead{3D-HST$/$CANDELS} & \colhead{Right Ascension} & \colhead{Declination} & \colhead{$m_{F160W}$} & \colhead{$z_{phot}$} & \colhead{$z_{grism}$} & \colhead{Notes} \\
 &  Catalog ID & (hours) & (degrees) &  &  &  & }
\startdata
      3 &        GOODS-S-41192 &         03:32:14.072 &         -27:43:03.72 &      23.54 &   3.206$_{-  0.058}^{+  0.065}$ &                3.153$_{-0.004}^{+0.019}$ &                                                     \\ 
      8 &        GOODS-S-41189 &         03:32:14.886 &         -27:43:03.54 &      24.83 &   3.246$_{-  0.101}^{+  0.104}$ &      -  &                                                     \\ 
      9 &        GOODS-S-41027 &         03:32:14.070 &         -27:43:05.27 &      26.06 &   3.179$_{-  0.171}^{+  0.160}$ &      -  &                                                     \\ 
     10 &        GOODS-S-41047 &         03:32:13.947 &         -27:43:04.81 &      26.02 &   3.204$_{-  0.293}^{+  0.308}$ &      -  &                                                     \\ 
     11 &        GOODS-S-41395 &         03:32:14.550 &         -27:42:59.34 &      25.36 &   3.036$_{-  0.085}^{+  0.085}$ &      -  &                                                     \\ 
     12 &        GOODS-S-41673 &         03:32:14.884 &         -27:42:54.27 &      27.26 &   3.231$_{-  0.212}^{+  0.188}$ &      -  &                                                     \\ 
 & & & & & \\ 
\hline
 & & & & & \\ 
      6 &                      &         03:32:14.606 &         -27:43:06.14 &            &   3.801$_{-  0.513}^{+  1.786}$ &       - &                      IRAC data / SF templates only  \\ 
        &                      &                      &                      &            &        $_{       }^{       }$ &         &           included in the photometric redshift fit  \\ 
        &                      &                      &                      &            &   3.094$_{-  0.272}^{+  2.097}$ &       - &          IRAC and MIPS data / AGN and SF templates  \\ 
        &                      &                      &                      &            &        $_{       }^{       }$ &         &               included in photometric redshift fit  \\ 
\enddata
\tablenotetext{a}{The ID column is based on the numbering scheme used in \citet{nil06}.}
\label{tab:galmember}
\end{deluxetable}
\end{turnpage}

\clearpage 


\clearpage 

\clearpage

\clearpage

\clearpage

\clearpage

\clearpage

\clearpage

\clearpage 

\clearpage

\clearpage

\clearpage

\clearpage

\end{document}